\documentclass[aps,prd,nofootinbib,showpacs,superscriptaddress]{revtex4} 
\usepackage[sumlimits]{amsmath}
\usepackage{epsfig}
\usepackage{amssymb}
\usepackage{graphicx}
\usepackage{pstricks}
\usepackage{color}
\usepackage{bbold}
\usepackage{hyperref}
\usepackage{cancel}
\usepackage{ragged2e}
\usepackage[font=small,labelfont=bf,justification=raggedright]{caption}

\newcommand{\be}{\begin{equation}}
\newcommand{\ee}{\end{equation}}
\newcommand{\bea}{\begin{eqnarray}}
\newcommand{\eea}{\end{eqnarray}}

\newsavebox{\measurebox}

\begin{document}

\title{Warm inflation within a supersymmetric distributed mass model}
\author{Mar Bastero-Gil}
\email{mbg@ugr.es}
\affiliation{Departamento de F\'{i}sica T\'{e}orica y del Cosmos, Universidad de Granada, Granada-18071, Spain}
\author{Arjun Berera}
\email{ab@ph.ed.ac.uk}
\author{Rafael Hern\'{a}ndez-Jim\'{e}nez}
\email{s1367850@sms.ed.ac.uk}
\affiliation{School of Physics and Astronomy, University of Edinburgh, Edinburgh, EH9 3FD, United Kingdom}
\author{Jo\~{a}o G. Rosa}
\email{joao.rosa@ua.pt}
\affiliation{Departamento de F\'{i}sica da Universidade de Aveiro and CIDMA, Campus de Santiago, 3810-183 Aveiro, Portugal}

\begin{abstract}
We study the dynamics and observational predictions of warm inflation within 
a supersymmetric distributed mass model. This dissipative mechanism is well 
described by the interactions between the inflaton and a tower of chiral 
multiplets with a mass gap, such that different bosonic and fermionic fields 
become light as the inflaton scans the tower during inflation.
We examine inflation for various mass distributions,
analyzing in detail the dynamics 
and observational predictions.
We show, in particular, that warm inflation can be consistently realized 
in this scenario for a broad parametric range and in excellent agreement 
with the Planck legacy data. Distributed mass models can be viewed
as realizations
of the landscape property of string theory, with the mass distributions
coming from the underlying spectra of the theory, which themselves
would be affected by
the vacuum of the theory. 
We discuss the recently proposed swampland criteria for inflation models
on the landscape and analyze 
the conditions under which they can be met within the distributed mass 
warm inflation scenario.  We demonstrate mass distribution models
with a range of consistency with the swampland criteria 
including cases in excellent consistency.
\end{abstract}
 \maketitle


\section{Introduction}

The most recent cosmological observations once again confirm an expanding 
universe that is spatially flat, homogeneous and isotropic on large scales, 
and where the large scale structure originated from primordial fluctuations 
with a nearly scale-invariant, adiabatic, and gaussian 
spectrum \cite{Planck}.  Inflation \cite{inflation} remains the 
dominant paradigm that can consistently explain the observational
data. In the standard inflation picture, cold inflation (CI), 
depicted by a homogeneous scalar ``inflaton''
field, the short period of quasi-de Sitter accelerated expansion 
phase quickly dilutes away all traces of any pre-inflationary matter 
or radiation density, so that the state of the universe is the vacuum 
state. However, this generates a supercooled universe and leaves 
indeterminate a reasonable description of the transition from inflation 
to the ``hot Big Bang" scenario, required by Big Bang Nucleosynthesis, 
and the physics of recombination leading to the Cosmic Microwave Background 
(CMB) that we observe today. This necessarily requires the conversion 
of inflaton energy density into ordinary matter and radiation and thus 
to its interactions with other fields. 

In the conventional inflation picture the inflaton decay can only play 
a significant role at the end of the slow-roll regime, since 
particle production is not pictured to occur within the inflationary expansion 
phase.  This leads to cold inflation ending through
the standard ``(p)reheating" paradigm \cite{reheating}. 
The reasoning behind this phase lies in the fact that the perturbative 
decay width of 
a particle is generically smaller than its mass, which in turn lies 
below the Hubble expansion rate for a slowly rolling scalar field. 
Consequently such interplay between the inflaton and other constituents 
may perform a negligible role during the slow-roll phase 
of inflationary models. Nonetheless, it is relevant to note that 
the perturbative decay width only describes the decay of a field close 
to the minimum of its potential \cite{Graham:2008vu}, which is evidently 
not the case during slow-roll dynamics, and that finite temperature 
effects can further significantly enhance the rate at which the inflaton 
dissipates its energy into other degrees of freedom. Thereby the inflaton 
field could be coupled to other components and might dissipate its 
vacuum energy and warm up the universe. This alternative scenario is 
known as the \emph{warm inflation} (WI) 
paradigm \cite{Berera:1995wh,Berera:1995ie}, where dissipative effects 
and associated particle production can, in fact, sustain a thermal bath 
concurrently with the accelerated expansion of the universe during inflation. 

One of the earliest warm inflation models \cite{Berera:1996nv} suggested 
the idea that parameters in an inflation model could be randomly distributed. 
The distributed-mass-model (DM model) 
\cite{Berera:1998px,Berera:1999px,Berera:1999ws,Berera:1999wt} was subsequently 
proposed and built on this idea in the context of string theory. 
It observed \cite{Berera:1999wt} that models from string theory have 
states at many energy levels and the distribution of these levels 
is ultimately dictated by the string vacuum. In particular, depending on the 
details of the compactification, the state of Kaluza-Klein modes, 
and patterns of symmetry breaking, it will induce splittings of string 
energy levels. Thus, based on the specific properties of a given string 
realization, the string states will be distributed differently.  
The idea of such distribution of states can find a more dynamical 
explanation in the context of the landscape picture, whereby different 
ground states of string theory will in turn imply changes to the 
spectrum of string states.  In this respect, the DM model was one of 
the first low-energy realizations following the 
landscape idea well before the idea 
was formally stated.

The string landscape idea \cite{Douglas:Caltech,Douglas:2003um}
claims an unimaginable huge number 
of possible vacua reaching by some estimates to
order $10^{500}$.  To have any hope in dealing with such
a large number of possibilities, one would need to 
work both from the direction of string theory and phenomenology to 
identify viable vacua.  The string landscape emerges from the complex
structure of string theory.  One property of this complex structure
is the huge number of string states, going into the hundreds of thousands.
Inflation models motivated by the landscape are generally built with only
a small number of fields and generally do not attempt to utilize the vast
spectra of string states.  DM models are one of the few that
attempt to utilize as part of the dynamical solution
this inherent feature that there are many string states available.
In the landscape
way of thinking therefore, DM models, although phenomenologically
motivated, presumably are studying 
theories emerging from a vast range of vacua that otherwise are
missed in inflation models with just a small number of fields.
As such, in constructing DM models it is natural 
to look for distributions that are of observational interest.  In this 
paper we will first look at the generic DM distribution, where the string 
states in the range relevant to inflation are equally spaced apart, 
which was the original example studied in the early 
work \cite{Berera:1998px,Berera:1999px,Berera:1999ws,Berera:1999wt}.  
This case has many 
relevant features but we find that it is not consistent with present 
observational constraints. We will then examine other types of of 
DM distribution that have appealing consequences in comparison with observation.  

Inflation is assumed to be described by low-energy Effective Field 
Theory (EFT), although in many models inflation can happens 
when the inflaton field is 
super-Planckian, particularly by considering monomial chaotic potentials
in cold inflation.
Furthermore, an EFT can be ultraviolet (UV) complete if it can be 
successfully incorporated in a quantum theory of gravity, such as 
String Theory. This paradigm provides a vast landscape of consistent 
embeddings of the EFT of gravity into a quantum theory, but this does not 
imply that any EFT coupled to gravity is consequently included in the 
landscapes. Those EFT's that are in fact inconsistent with a quantum 
theory of gravity lie in the surrounding 
swamplands \cite{Vafa:2005ui,Ooguri:2006in}. Hence, a benchmark is needed 
to ensure a de Sitter vacuum EFT can live in the desired string landscapes. 
Recently two swampland criteria relevant for inflationary theories have 
been proposed 
\cite{Obied:2018sgi,Agrawal:2018own}, $|\Delta\phi|/M_{P}<\Delta$ and $M_{P}|V_{\phi}|/V>c$, provided that $V>0$, 
where $\{\Delta,c\}\sim\mathcal{O}(1)$. Nonetheless, these criteria have been 
noted to pose inherent threats to the basic mechanism of slow-roll in 
cold inflation \cite{Agrawal:2018own}. However, as part of the subsequent 
analysis we evaluate the aforementioned criteria 
in the warm inflation scenario.  We show models of warm inflation
that can very
consistent with the swampland criteria.  Inherently the dissipative
feature in warm inflation makes it amenable for consistency
with swampland criteria, as already noted in the
literature \cite{Das:2018hqy,Das:2018rpg,Motaharfar:2018zyb,Yi:2018dhl,Lin:2018edm}.

In this paper we will examined in detail DM models and explore
their consistency with observational data and theoretical viability.
This work is organized as follows. 
In Sect.~\ref{WARM INFLATION DYNAMICS AND PRIMORDIAL PERTURBATION SPECTRUM} 
we introduce all WI dynamics and primordial perturbation spectrum.
In Sect.~\ref{Supersymmetric distributed mass model} we present in detail 
the supersymmetric distributed mass model, which is described by the 
interactions between the inflaton field and other light constituents: 
fermion ($\{\psi_{i}\,,\psi_{\sigma}\}$) and 
scalar ($\{\chi_{i}\,,\sigma\}$) fields. In Sect.~\ref{interactions} we 
calculate all relevant parameters of the dissipation dynamics for the 
bosonic and fermionic sector: dissipative coefficients and their 
corresponding thermal average of the decay width. 
In Sect.~\ref{Distinct mass site configurations} we develop
a form of the mass distribution function.
In Sect.~\ref{Results} we analyze in detail various 
dissipative coefficients for 
several inflation driven monomial potentials; we apply 
standard slow-roll methods and identify
observationally consistent regions in parameter space. Here we also
test the recently proposed swampland criteria. 
The main conclusions of this work are summarized in 
Sect.\ref{Conclusions}. Three appendices are also included, where we provide 
more detailed discussions of some of the results used in our computations. 


\section{WARM INFLATION DYNAMICS AND PRIMORDIAL PERTURBATION SPECTRUM}\label{WARM INFLATION DYNAMICS AND PRIMORDIAL PERTURBATION SPECTRUM}

Non-equilibrium effects in the dynamics of a scalar field are generically produced due to interactions with an ambient thermal bath. The leading non-equilibrium effect, for a field evolving slowly compared to the characteristic time scale of the thermal bath, is a dissipative friction term $\Upsilon\dot\phi$ in its equation of motion \cite{Berera:2008ar}, where $\Upsilon=\Upsilon(\phi,T)$ can be computed from first principles given the form of the interactions between the scalar field and the thermalized degrees of freedom. For a homogeneous field, this implies the continuity equation $\dot\rho_\phi + 3H(\rho_\phi+p_\phi)=-\Upsilon\dot\phi^2$, such that overall energy-momentum conservation implies the existence of an identical term with opposite sign in the continuity equation for the thermal fluid. This explicitly proves that dissipative effects in the inflaton's equation of motion lead to particle production in the thermal bath, so it prevents the exponential dilution of the latter in a quasi-de Sitter background. Therefore the temperature does not drop abruptly and the universe is able to smoothly cross to the radiation epoch. Taking into account dissipative effects, the evolution equation for the background inflaton field is given by:
\begin{equation}
\ddot{\phi}+\left(3H+\Upsilon\right)\dot{\phi}+V_{eff,\phi}(\phi,T)=0 \,,
\end{equation}
where $V_{eff,\phi}=dV_{eff}/d\phi=V(\phi)_{,\phi}+V_{T,\phi}$ includes the effective thermal potential $V_{eff}=V(\phi)+V_{T}$, where $V(\phi)$ is the zero temperature potential, and $V_{T}$ is the finite temperature effective potential, including radiative corrections to the potential due to any light component (fermion or scalar fields); $\Upsilon$ is the dissipative coefficient and $H$ is the Hubble parameter, given by the Friedmann equation for a flat FRW universe:
\begin{equation}\label{Hubble}
H^{2}=\frac{\rho_{T}}{3M_{P}^{2}}\,, 
\end{equation}
where $\rho_{T}=\dot{\phi}^{2}/2+V_{eff}+Ts_{R} $ is the total energy density of the system, $s_{R}$ is the entropy density, and and $M_P$ is the reduced Planck mass. One parameter that quantifies dissipation during WI is defined as the dissipative ratio $Q=\Upsilon/(3H)$. Depending on the ratio $Q$ we can have different regimes: when $Q < 1$, this is called weak dissipative warm inflation; and when $Q\gtrsim 1$ we are in strong dissipative warm inflation. Furthermore,  the WI paradigm assumes the presence of a thermal bath, at temperature $T$; hence the evolution of such dissipative mechanism can be obtained from the evolution equation for the entropy density, given by:
\begin{equation}\label{entropy-evolution}
T\left(\dot{s}_{R}+3Hs_{R} \right) = \Upsilon \dot{\phi}^{2} \,,
\end{equation}
which in the slow-roll regime reduces to $3HTs_{R} \simeq \Upsilon \dot{\phi}^{2}$. Without including the $T$-dependent corrections in the inflaton potential, we would have the standard relation $Ts_{R}=4\rho_{R}/3=4C_{R}T^{4}/3$, where $C_{R}=\pi^{2}g_{eff}(T)/30$, $g_{eff}(T)$ is the effective $T$-dependent number of relativistic degrees of freedom, with $\rho_{R}$ denoting the standard energy density of radiation. All relativistic light fields contribute to the effective degrees of freedom, having:
\begin{equation}\label{geffT}
g_{eff}(T)=\frac{90}{4\pi^{2}}\frac{s_{R}}{T^{3}}\,. 
\end{equation}

Since the radiative corrections modify the potential and its derivatives, they alter the standard slow-roll parameters, being from $\epsilon_{\phi}=M_{P}^{2}(V_{,\phi}/V)^{2}/2$ and $\eta_{\phi}=M_{P}^{2}V_{,\phi\phi}/V$ to: 
\begin{equation}\label{effective-slow-roll}
\epsilon_{eff}=\frac{M_{p}^{2}}{2}\left(\frac{V_{eff,\phi}}{V_{eff}}\right)^{2} \,,\qquad \eta_{eff}=M_{p}^{2}\frac{V_{eff,\phi\phi}}{V_{eff}} \,, 
\end{equation}
where $V_{eff,\phi\phi}=V(\phi)_{,\phi\phi}+V_{T,\phi\phi}$. Recall that inflation happens when $\dot{\phi}^{2}/2\ll V_{eff}$ so does $(Ts_{R}Q^{-1}/2)\ll V_{eff}$, but even if small compared to the inflaton effective potential, it can be larger than the expansion rate with $(Ts_{R})^{1/4}\gtrsim H$; by assuming thermalization, this translates roughly into $T\gtrsim H$, so one can obtain a warm inflationary universe consistently with a slow-roll evolution, as long as the radiative corrections are restrained such they do not spoil inflation. Furthermore, one can show that the radiation energy density portrayed by an entropic description can never exceed the inflationary potential in a slow-roll regime, guaranteeing a period of accelerated expansion: 
\begin{equation}\label{radiation_abundance}
\frac{Ts_{R}}{V_{eff}}\simeq \frac{2}{3}\frac{\epsilon_{eff}}{1+Q}\frac{Q}{1+Q} \,,
\end{equation}
such that consistency of the slow-roll evolution 
requires $\epsilon_{eff}<1+Q$. This in turn also implies that, 
at the end of the slow-roll regime, when $\epsilon_{eff}\sim 1+Q$, one 
may attain $Ts_{R}\sim V_{eff}$ if a strong dissipative 
regime $Q\gtrsim 1$ can be achieved. In such cases radiation will 
smoothly become the dominant component at the end of inflation, 
providing the necessary ``graceful exit" into the ``hot Big Bang" cosmic 
evolution \cite{Berera:1996fm}. Although there may be additional particle 
production at the end of inflation, no reheating is actually necessary 
in WI when strong dissipation is reached, otherwise such mechanism is 
needed. Additionally to the smooth exit from inflation, WI exhibits 
several attractive features that have been explored in recent years. 
For instance, the dissipative friction damps the inflaton's evolution, 
making slow-roll easier or, equivalently, alleviating the conditions 
on the flatness of the inflaton potential, expressed now by the 
slow-roll conditions $\epsilon_{eff}, |\eta_{eff}|\ll 1+Q$. This may 
potentially provide a solution to the so-called ``eta-problem" typically 
found in string/supergravity inflationary models 
where generically 
$\eta_\phi \sim \mathcal{O}(1)$ \cite{Berera:1999ws,BasteroGil:2009ec}.
(For other recent reviews of warm inflation, please see
\cite{Oyvind Gron:2016zhz,Rangarajan:2018tte}.)

Small fluctuations of the inflaton about its homogenous component provide the initial seeds of density perturbation. These density perturbations produced during inflation evolve into the classical inhomogeneities observed in the CMB. For WI scenarios the fluctuations of the inflaton are thermally induced. As such, these initial seeds of density perturbations are already classical upon definition. The general expression for the amplitude of the primordial spectrum is given by \cite{Hall:2003zp, Moss:2007cv, Graham:2008vu, Ramos:2013nsa, BasteroGil:2011xd}:
\begin{equation}\label{general_spectrum}
\Delta_{\mathcal{R}}^{2} = \left(\frac{H_{*}}{\dot{\phi}_{*}}\right)^{2}\left(\frac{H_{*}}{2\pi}\right)^{2}\left(1+2n_{*}+\frac{2\sqrt{3}\pi Q_{*}}{\sqrt{3+4\pi Q_{*}}}\frac{T}{H}\right)G(Q_{*}) \,,
\end{equation} 
where all quantities are evaluated when the relevant CMB modes become 
superhorizon 50-60 e-folds before inflation ends. In the expression 
above, $n_{*}$ denotes the inflaton phase space distribution at horizon 
crossing. By the strength of the interactions between the inflaton field 
and other particles in the thermal bath (including e.g. scattering 
processes), this might interpolate between the Bunch-Davies 
vacuum, $n_{*}=0$, and the Bose-Einstein distribution at the ambient 
temperature $T$, $n_{*}\simeq\left(e^{H_{*}/T_{*}}-1  \right)^{-1}$. We 
will focus on the latter limiting case in this paper, which we denote 
as ``thermal" inflaton fluctuations. Also the function $G(Q_{*})$ accounts 
for the growth of inflaton fluctuations due to the coupling to radiation 
fluctuations through the temperature dependence of the dissipation 
coefficient and must be determined numerically. In addition, this function 
also exhibits a mild dependence on the form of the scalar potential. 
In general, with thermalised inflation 
fluctuations $1+2n_{*}=\coth\left(H_{*}/(2T_{*})\right)$, we have:
\begin{equation}\label{general_thermal-spectrum}
\Delta_{\mathcal{R}}^{2} \simeq \left(\frac{3H_{*}^{3}(1+Q_{*})}{2\pi V_{eff,\phi\,*}}\right)^{2}\left(\frac{2\sqrt{3}\pi Q_{*}}{\sqrt{3+4\pi Q_{*}}}\frac{T_{*}}{H_{*}}+\coth\left(\frac{H_{*}}{2T_{*}}\right) \right)G(Q_{*}) \,,
\end{equation} 
then from the amplitude of the curvature power spectrum, we may determine the scalar spectral index $n_{s}-1\simeq d\ln\Delta_{\mathcal{R}}^{2}/dN_{e}$. Since, for $T\ll M_P$, gravitational waves are not significantly affected by thermal effects, the primordial tensor spectrum is given by the standard inflationary form $\Delta_{t}^{2}=2H_{*}^{2}/(\pi^{2}M_P^{2})$. The tensor-to-scalar ratio $r=\Delta_{t}^{2}/\Delta_{\mathcal{R}}^{2}$ is nevertheless affected, and in fact typically reduced, by the modifications to the scalar curvature perturbations introduced due to dissipation, which are basically a function of $T_{*}/H_{*}$ and $Q_{*}$. We illustrate this fact by using the slow-roll dynamics, where the ratio $r$ can be written as:
\begin{equation}\label{SR-r}
r \simeq \frac{16\epsilon_{eff}}{(1+Q_{*})^{2}F(T_{*}/H_{*},Q_{*})} \,,\quad F(T_{*}/H_{*},Q_{*})=\left(\frac{2\sqrt{3}\pi Q_{*}}{\sqrt{3+4\pi Q_{*}}}\frac{T_{*}}{H_{*}}+\coth\left(\frac{H_{*}}{2T_{*}}\right) \right)G(Q_{*}) \,,
\end{equation}
which is suppressed w.r.t. the CI prediction by a factor $(1+Q_{*})^{2}K(T_{*}/H_{*},Q_{*})>1$. Indeed in \cite{BasteroGil:2009ec} was shown explicitly, even before the BICEP and Planck results, that the presence of radiation and dissipation suppresses the tensor-to-scalar ratio. Authors in \cite{BasteroGil:2009ec} computed the tensor-to-scalar ratio of the monomial $\phi^2$ and $\phi^4$ models and was one of the few analyses at the time that predicted for these inflation driven potentials a low tensor-to-scalar ratio, which now we see is consistent with data. Subsequent work further developed the analysis \cite{Cai:2010wt, Bartrum:2013fia, Bastero-Gil:2016qru, Bastero-Gil:2017wwl, Bastero-Gil:2018uep}, showing that the tensor-to-scalar ratio may attain values even below $10^{-3}$ for a $\phi^4$ potential, and thus potentially distnguishable from e.g.~scenarios with a non-minimal coupling to gravity such as Higgs inflation \cite{Bezrukov:2007ep} or the Starobinsky model \cite{inflation}.



\section{Supersymmetric distributed mass model}
\label{Supersymmetric distributed mass model}

Let us consider the general form of an effective $\rm N=1$ global SUSY theory version of the distributed mass (DM) model with chiral superfields $\Phi$, $X_{i}$ and $Y_{i}$, described by the superpotential \cite{Hall:2004zr,BasteroGil:2009ec,BasteroGil:2010pb,BasteroGil:2012cm}:
\begin{equation}\label{superpotential}
W=\sum_{i}\left[\frac{g}{2}(\Phi-M_{i}) X_{i}^{2}+\frac{h}{2}X_{i}Y_{i}^{2} \right]\,,
\end{equation}  
where $g$ and $h$ are coupling constants and the sum is taken over an arbitrary distribution of supermultiplets $X_i$ and $Y_i$. The chiral superfields $\Phi$, $X_{i}$ and $Y_{i}$ have (scalar, fermion) components ($\phi$,$\psi_{\phi}$), ($\chi_{i}$,$\psi_{\chi_{i}}$) and ($\sigma_{i}$,$\psi_{\sigma_{i}}$), respectively. Note that these are complex scalars and Weyl fermions, each with two degrees of freedom. We may use the Majorana representation for the spinors, i.e. use a 4-component Majorana spinor built from the same Weyl fermion. In this case, note that a Majorana fermion is its own anti-particle. We have considered different $Y_{i}$ fields coupled to each $X_{i}$ field in the tower to avoid mass mixing at the level of the thermal masses. The scalar interaction terms in the theory are obtained from the superpotential using \cite{Hall:2004zr}: 
\begin{equation}\label{lagrangian-LS-definintion}
-\mathcal{L}_{S}=|\partial_{\Phi}W|^{2}+\sum_{i}|\partial_{X_{i}}W|^{2}+\sum_{i}|\partial_{Y_{i}}W|^{2}\,.
\end{equation} 
On the other hand the fermion Lagrangian can be computed from the general formula \cite{Hall:2004zr}:
\begin{equation}\label{lagrangian-LF-definition}
-\mathcal{L}_{F}=\frac{1}{2}\sum_{n,m}\frac{\partial^{2}W}{\partial\xi_{n}\partial\xi_{m}}\bar{\psi}_{n}P_{L}\psi_{m}+\frac{1}{2}\sum_{n,m}\frac{\partial^{2}W^{\dagger}}{\partial\xi^{\dagger}_{n}\partial\xi^{\dagger}_{m}}\bar{\psi}_{n}P_{R}\psi_{m}\,,
\end{equation}
where $\xi_{n}$ is a superfield: $\Phi,X_{i}, Y_{i}$, and $P_{L}=1-P_{R}=(1+\gamma_{5})/2$ are the chiral projection operators acting on Majorana 4-spinors. Note that $\partial W/\partial X_{i}\partial Y_{i}=\partial W/\partial Y_{i}\partial X_{i}$ and similarly $\partial W^{\dagger}/\partial X^{\dagger}_{i}\partial Y^{\dagger}_{i}=\partial W^{\dagger}/\partial Y^{\dagger}_{i}\partial X^{\dagger}_{i}$. Then we select only the boson field components of the chiral superfields $\Phi$, $X_{i}$ and $Y_{i}$, which are $\phi$, $\chi_{i}$ and $\sigma_{i}$, respectively. Nonetheless, see that the inflaton field $\phi$ can be decomposed into its real and imaginary parts via\footnote{Similarly we define the complex scalars in terms of the real and imaginary part for $\chi_{i}$ and $\sigma_{i}$.} $\phi=(\phi_{R}+i\phi_{I})/\sqrt{2}$. Hence, this prescription introduces another decay channel; for instance, the modulus square becomes $|\phi-M_{i}|^2=(\phi_{R}/\sqrt{2}-M_{i})^{2}+\phi_{I}^{2}/2$. However, $\phi_{R}/\sqrt{2}$ is the nonzero vacuum expectation value of $\phi$, which will be the only term that will contribute to dissipation in the scalar field's effective equation of motion, so that the imaginary part of the inflaton, $\phi_{I}$, is not relevant for our subsequent analysis. Moreover, in order to be coherent with further calculations, $\phi$ is going to be considered only as the classical expectation value, without the label $R$ and the factor $1/\sqrt{2}$. Therefore, the relevant Lagrangians that may contribute to dissipation in the scalar field's effective equation of motion are: 
\begin{eqnarray}\label{lagrangians-full2}
-\mathcal{L}_{S} &=& g^{2}\sum_{i}\left(\phi-M_{i}\right)^{2}|\chi_{i}|^{2}+\frac{gh}{2}\sum_{i}\left(\phi-M_{i}\right)\left[\chi_{i}(\sigma_{i}^{\dagger})^{2}+\chi_{i}^{\dagger}\sigma_{i}^{2}\right] \nonumber\\
&& + h^{2}\sum_{i}|\chi_{i}|^{2}|\sigma_{i}|^{2}+ \frac{g^{2}}{4}\sum_{i}|\chi_{i}|^{4} + \frac{h^{2}}{4}\sum_{i}|\sigma_{i}|^{4} \,, \\
-\mathcal{L}_{F} &=& \frac{g}{2}\sum_{i}\left(\phi-M_{i}\right)\bar{\psi}_{\chi_{i}}P_{L}\psi_{\chi_{i}} + \frac{g}{2}\sum_{i}\left(\phi-M_{i}\right)\bar{\psi}_{\chi_{i}}P_{R}\psi_{\chi_{i}} \nonumber\\
&& + \frac{h}{2}\sum_{i}\chi_{i}\bar{\psi}_{\sigma_{i}}P_{L}\psi_{\sigma_{i}} + \frac{h}{2}\sum_{i}\chi^{\dagger}_{i}\bar{\psi}_{\sigma_{i}}P_{R}\psi_{\sigma_{i}} + h\sum_{i}\sigma_{i}\bar{\psi}_{\sigma_{i}}P_{L}\psi_{\chi_{i}} + h\sum_{i}\sigma^{\dagger}_{i}\bar{\psi}_{\sigma_{i}}P_{R}\psi_{\chi_{i}}\,. 
\end{eqnarray}
Note that the bare masses are $m_{\psi_{\chi_{i}}}=g\left(\phi-M_{i}\right)=m_{\chi_{i}}$ at zero temperature for unbroken SUSY\footnote{Note that in warm inflation SUSY is broken both by the finite temperature and the inflaton energy density. The latter should arise from an additional $\Phi$-dependent term in the superpotential that we have note included above and that will lead to a small splitting of the mass for the real and imaginary components of the $\chi_{i}$ scalar fields.}. This is in agreement with \cite{Hall:2004zr} upon rescaling the couplings $g$ and $h$ in the superpotential by $1/2$ factors.

At finite temperature, both the $\chi_{i}$ and $\psi_{\chi_{i}}$ receive thermal mass corrections. The contributions of the $\sigma_{i}$ and $\psi_{\sigma_{i}}$ fields to the latter have been computed in \cite{Hall:2004zr} and been shown to be identical for both $\chi_{i}$ and $\psi_{\chi_{i}}$, corresponding to $h^{2}T^{2}/8$ taking into account the coupling normalization differences. The $\chi_{i}$ scalars also receive thermal corrections from their self-interactions $g^{2}|\chi_{i}|^{4}/4$. Noting that these interactions give a contribution to their tree-level mass $\partial^{2}V/\partial\chi_{i}\partial\chi_{i}^{\dagger}= g^{2}|\chi_{i}|^{2}$ and taking into account the contribution of the $\chi_{i}$ fields to the thermal effective potential:
\begin{equation}
\Delta V_{T}\subset 2\times\frac{m_{\chi_{i}}^{2}}{24}T^{2}=\frac{g^{2}}{12}T^{2}|\chi_{i}|^{2}+\ldots \,,
\end{equation}
where we have taken into account the two degrees of freedom for complex scalars, this yields a thermal mass correction $g^{2}T^{2}/12$ to the $\chi_{i}$ fields\footnote{The mass of the scalar $\sigma_i$ fields is also corrected by a similar factor (involving only the $h$ coupling) and, although the resulting masses are below the temperature, they will generically lie above $H$ in the parametric range relevant to our discussion. This implies that no scalar field other than the inflaton can sustain a slowly-rolling background value that drives inflation.}. In summary, we obtain:
\begin{equation}
\Delta m_{\chi_{i}}^{2}=\frac{g^{2}}{12}T^{2}+\frac{h^{2}}{8}T^{2}\,,\qquad \Delta m_{\psi_{\chi_{i}}}^{2}=\frac{h^{2}}{8}T^{2} \,.
\end{equation}

We also need to compute the finite temperature decay widths of the $\chi_{i}$ and $\psi_{\chi_{i}}$ fields. However, we only recall such calculations for Dirac fermions \cite{BasteroGil:2010pb}; albeit the difference between Majorana and Dirac fermions is only in the overall factors of the decay width. Hence, we can first compute them at zero temperature to set the correct normalisation factors in order to identify these global constants. In general we have for the decay of a particle of mass $m$ at rest into a pair of massless particles:
\begin{equation}
\Gamma=\frac{S}{16\pi m}|\mathcal{M}|^{2} \,,
\end{equation}
where $S=1/2$ if the particles are identical and $S=1$ if they are distinct.
The $\chi_{i}$ fields may decay via $\chi_{i}\rightarrow\sigma_{i}\sigma_{i}$ and $\chi_{i}\rightarrow\psi_{\sigma_{i}}\psi_{\sigma_{i}}$. In the first case, dropping the indices for simplicity, we may decompose the fields into their real and imaginary components via $\chi=(\chi_{R}+i\chi_{I})/\sqrt{2}$ and analogously for $\sigma$. This yields the scalar interactions:
\begin{equation}
-\mathcal{L}_{\chi\sigma^{2}}=\frac{hg}{2\sqrt{2}}(\phi-M_{i})\left[\chi_{R}\sigma_{R}^{2}-\chi_{R}\sigma_{I}^{2}+2\chi_{I}\sigma_{R}\sigma_{I}\right]\,.
\end{equation}
Hence, the $\chi_{R}$ scalar may decay into $\sigma_{R}$ or $\sigma_{I}$ pairs, while the $\chi_{I}$ scalar has only one decay channel $\chi_{I}\rightarrow\sigma_{R}\sigma_{I}$. For each of these decay channels, the vertex factor is $-ihg(\phi-M_{i})/\sqrt{2}$. Taking into account the $S=1/2$ factors in the $\chi_{R}\rightarrow\sigma_{R}\sigma_{R}$ and $\chi_{R}\rightarrow\sigma_{I}\sigma_{I}$ channels, we then find that the decay widths are equal for both $\chi_{R}$ and $\chi_{I}$, being given by:
\begin{equation}
\Gamma_{\chi_{i}}^{S}=\frac{h^{2}g^{2}(\phi-M_{i})^{2}}{32\pi m_{\chi_{i}}} \,.
\end{equation}
The fermionic decay channel comes from the interaction term $\frac{1}{2}h\chi_{i}\bar{\psi}_{\sigma_{i}}\psi_{\sigma_{i}}$, where the vertex factor is simply $-ih$. Since the particles are identical in the final state and computing the matrix element with the usual Feynman rules, this yields:
\begin{equation}\label{gamma-F-chi-T-0}
\Gamma_{\chi_{i}}^{F}=\frac{h^{2}}{32\pi}m_{\chi_{i}} \,.
\end{equation}
Noting that, at zero temperature, $m_{\chi_{i}}=g(\phi-M_{i})$, we see that $\Gamma_{\chi_{i}}^{S}=\Gamma_{\chi_{i}}^{F}$ in this limit.

The fermions $\psi_{\chi_{i}}$ can decay as $\psi_{\chi_{i}}\rightarrow\sigma_{i}\psi_{\sigma_{i}}$ via the corresponding Yukawa term above, and this yields simply:
\begin{equation}\label{gamma-F-psi-T-0}
\Gamma_{\psi_{\chi_{i}}}=\frac{h^{2}}{16\pi}m_{\psi_{\chi_{i}}} \,.
\end{equation}
Again, note that at zero temperature for unbroken SUSY we have $m_{\chi_{i}}=m_{\psi_{\chi_{i}}}=g(\phi-M_{i})$, which yields identical total decay widths for the scalars and fermionic superpartners, as it should.

Once we have compute the decay widths at zero temperature, we identify each vertex factor squared from the decay of a scalar boson to fermions and the decay of a fermion to a scalar boson and a fermion. In the pole approximation (see Appendix \ref{appendix a}) we take the limit $T\rightarrow 0$ of the corresponding decay widths \cite{vertex-factors} and then by comparing them with eqs.~(\ref{gamma-F-chi-T-0}) and (\ref{gamma-F-psi-T-0}), we identify: $g_{\chi_{i}\bar{\psi}_{\sigma_{i}}\psi_{\sigma_{i}}}^{2}=h^{2}/4$ and $g_{\psi_{\chi_{i}}\sigma_{i}\psi_{\sigma_{i}}}^{2}=h^{2}$. From here we will use these overall factors for the rest of the computations.


\section{Dissipative interactions}\label{interactions}

The particle physics model considered in this work is inspired by string theory exhibiting $\mathcal{N}=1$ global supersymmetry, with the inflaton field coupled to massive modes of the string, as discussed in \cite{Berera:1998cq}. Several interactions are identified by the shifted couplings $g^{2}(\phi-M_{i})^{2}\chi_{i}^{2}$, and $g(\phi-M_{i})\bar{\psi}_{\chi_{i}}\psi_{\chi_{i}}$ for bosons $\chi_{i}$ and fermions $\psi_{\chi_{i}}$ respectively, with $\{M_{i}\}$ ranging over mass scales. This feature yields the name of distributed-mass-model (DM-model). In the subsequent segments we establish the relevant interplay the inflaton field has with the aforementioned fields. We will restrict to interactions such that the leading contribution will come dominantly from one-loop processes, when the decaying field is light. The key property of the DM model is that for a given temperature $T$, only the fields with masses $g^{2}(\phi-M_{i})^{2}\lesssim T^{2}$ will contribute to the dissipation. Henceforth, we will refer such configuration of states as thermally excited sites. We will consider separately the dissipative processes associated with the excitation of the scalar $\chi_i$ fields, which may decay via $\chi_i\rightarrow \sigma_i\sigma_i$ or $\chi_i\rightarrow \bar{\psi}_{\sigma_i}\psi_{\sigma_i}$, and those associated with the excitation of the fermionic $\psi_{\chi_i}$ fields, which decay via $\psi_{\chi_i}\rightarrow \sigma_i\psi_{\sigma_i}$, with technical details of the computation given in Appendix \ref{appendix a}.

\subsection{Bosonic sector}

The dissipative coefficient arising from the pattern of interactions among the scalar component and the light states is given by \cite{Berera:1998px,Berera:1999px,Berera:1999ws}(see also Appendix \ref{appendix a}):
\begin{equation}\label{Upsilon_DMS}
  \Upsilon^{S}(\phi,T)= \sum_{i=1}^{t.e.}\frac{32 g^{4}}{\pi h^{2}\left[16\frac{m_{\chi_{i}}^{2}}{\tilde{m}_{\chi_{i}}^{2}}+\frac{\tilde{m}_{\chi i}^{2}}{T^{2}}\right]}\ln\left(\frac{2T}{\tilde{m}_{\chi_{i}}}\right)\frac{\left(\phi-M_{i}\right)^{2}}{\tilde{m}_{\chi_{i}}}  \,,
\end{equation}
where `t.e.' means sum over all thermally excited sites. In the computation of the dissipative coefficient, the total decay rate for all processes is already taken into account, which is given by \cite{Berera:1998px,Berera:1999px,Berera:1999ws}(see also Appendix \ref{appendix a}): 
\begin{equation}\label{Gamma_DMS}
\Gamma_{\chi_i}=\Gamma(\chi_i\rightarrow \sigma_i\sigma_i)+\Gamma(\chi_i\rightarrow \bar{\psi}_{\sigma_i}\psi_{\sigma_i}) =\frac{h^{2}}{128\pi}\frac{T\tilde{m}_{\chi_{i}}}{\omega_{\chi_{i}}(p)}\left[16\frac{m_{\chi_{i}}^{2}}{\tilde{m}_{\chi_{i}}^{2}}+\frac{\tilde{m}_{\chi i}^{2}}{T^{2}}\right] \,,
\end{equation}
where  $\omega_{\chi i}(p)=\sqrt{\tilde{m}_{\chi i}^{2}+|\mathbf{p}|^{2}}$ and $\tilde{m}_{\chi i}$ is the full $\chi_i$ mass, including the tree-level contribution $m_{\chi_i}$ and the thermal corrections computed above. We need to ensure that all light bosons remain in a nearly-thermal state during inflation, so that the above result for the dissipation coefficient is a consistent approximation. For simplicity, we may take the thermal average of the above decay width, such: 
\begin{equation}
\bar{\Gamma}_{\chi_i}=\frac{1}{n_{B}}\int\frac{d^{3}p}{(2\pi)^{3}}\Gamma_{\chi_i}f_{B} \,
\end{equation}
where $f_{B}(\mathbf{p}/T)$ is the Bose-Einstein distribution and $n_{B}(\tilde{m}_{\chi_{i}}/T)$ the associated number density. This yields in the limit $m_{\chi i}\lesssim T$ via a numerical approximation (see Appendix \ref{appendix b}):
\begin{equation}\label{decay-rate-approx_CDMS}
\bar{\Gamma}_{\chi_i}\simeq \frac{h^{2}f^{1/2}}{128\pi}\left[\frac{16}{f}\frac{m_{\chi_{i}}^{2}}{T^{2}}+f\right]\frac{0.68}{(1+0.77 f^{1/2})}T \,,
\end{equation}
where $f=f(g,h)=g^{2}/12+h^{2}/8$. Note that the above expression depends on the factor $m_{\chi_{i}}/T=g(\phi-M_i)/T$, so that different fields will decay at different rates. However, note that states for which $m_{\chi_i}/T=0$, corresponding to $\phi=M_i$, do not contribute to dissipation according to Eq.~(\ref{Upsilon_DMS}). Conversely, the heaviest states that can be thermally excited have $m_{\chi_i}\sim T$, and these are the ones that contribute the most to dissipation at any given time. Therefore,  in analyzing the consistency of the model, namely whether the $\chi_i$ decay faster than expansion in order to remain close to thermal equilibrium, we will consider the states for which $m_{\chi_{i}}/T\simeq 1$.

\subsection{Fermionic sector}

The calculation of the dissipation coefficient can be done following essentially the same steps as in the ``Warm Little Inflaton" model \cite{Bastero-Gil:2016qru,Bastero-Gil:2017wwl, Bastero-Gil:2018uep}, yielding:
\begin{equation}\label{Upsilon_DMF}
  \Upsilon^{F}(T)=\sum_{i=1}^{t.e.}C_{T}^{F}T \,, \quad C_{T}^{F} \simeq \frac{3g^{2}}{h^{2}(1-0.34\log(h))} \,.
\end{equation}
This calculation involves computing the finite temperature decay width of all interaction processes, which is given by
\begin{equation}\label{decay-rate-full}
\Gamma_{\psi_{\chi_i}}=\Gamma(\psi_{\chi_i}\rightarrow \sigma_i\psi_{\sigma_i})= \frac{h^{2}}{16\pi}\frac{T^{2}m_{\psi i}^{2}}{\omega_{\psi i}^{2}(p)|\mathbf{p}|}\left[F\left(\frac{k_{+}}{T},\frac{\omega_{\psi i}(p)}{T}\right)-F\left(\frac{k_{-}}{T},\frac{\omega_{\psi i}(p)}{T}\right) \right]  \,,
\end{equation}
where, neglecting the masses of the decay products $\sigma$ and $\psi_{\sigma}$, we have $\omega_{\psi i}(p)=\sqrt{\tilde{m}_{\psi i}^{2}+|\mathbf{p}|^{2}}$, $k_{\pm}=(\omega_{\psi i}(p)\pm |\mathbf{p}|)/2$ and
\begin{equation}
F(x,y)=xy-\frac{x^{2}}{2}+(y-x)\ln\left(\frac{1-e^{-x}}{1+e^{x-y}}\right)+\text{Li}_{2}\left(e^{-x}\right)+\text{Li}_{2}\left(-e^{x-y}\right) \,,
\end{equation}
where $\text{Li}_{2}(z)$ is the dilogarithm function. Note that the mass of the light fields is corrected by a factor $h^{2}T^{2}/8$ due to their interactions with the thermal bath fields $\psi_{\sigma_i}$ and $\sigma_i$, which we will take to be dominant over the inflaton contribution, i.e. we take $\tilde{m}_{\psi i}\simeq h^{2}T^{2}/8$ in the above computation of the decay width. We need to ensure that all light fermions remain in a nearly-thermal state during inflation, so that the above result for the dissipation coefficient is a consistent approximation. For simplicity, we may take the thermal average of the above decay width in the high-temperature and ultra-relativistic region:
\begin{equation}
\bar{\Gamma}_{\psi_{\chi_i}}=\frac{1}{n_{F}}\int\frac{d^{3}p}{(2\pi)^{3}}\Gamma_{\psi_{\chi_i}}f_{F} \,
\end{equation}
where $f_{F}(\mathbf{p}/T)$ is the Fermi-Dirac distribution and $n_{F}(\tilde{m}_{\chi_{i}}/T)$ the associated number density. This yields in the limit $m_{\psi_{\chi_i}}\lesssim T$:
\begin{equation}\label{decay-rate-approx_CDMF}
\bar{\Gamma}_{\psi_{\chi_i}}\simeq 10^{-3}\left[1-0.875\ln\left(\frac{h^{2}}{8}\right)\right]h^{4}T \,.
\end{equation}


\section{Mass distribution function}\label{Distinct mass site configurations}

In the DM model the dissipative mechanism is well described by the interaction 
between the inflaton $\phi$ and the finite number of fermion and boson 
fields that are light at any given time as the scalar field scans the 
tower of states.  DM models are distinguished by how the mass
sites are distributed.  Such an idea has a natural realization with
string theory, whereby the inflaton is suggestive of an excited string
zero mode, which then interacts with massive string levels. Such
a construction of DM models from string theory was 
shown in \cite{Berera:1999wt}.  String levels can be highly degenerate
and the distribution of mass states was suggested in \cite{Berera:1999wt}
to emerge as a fine structure splitting of such levels.
Thus the pattern of splitting of a level will depend on
the properties of the string state, ultimately governed by
the underlying string vacuum.  In this respect different string vacua
should imply different distribution of mass states.  A
first principles determination of such distributions is
difficult and minimally requires a detailed study of string theory
case-by-case for each different possible vacua.  However,
we can adopt a phenomenological approach and look at various
types of mass distributions and see what type of inflation
they can lead to.  This would be a minimum first step to test
the viability of such models.  In this section we develop
some basic properties of the mass distribution function.

\begin{figure}[h]
\centering\includegraphics[scale=0.9]{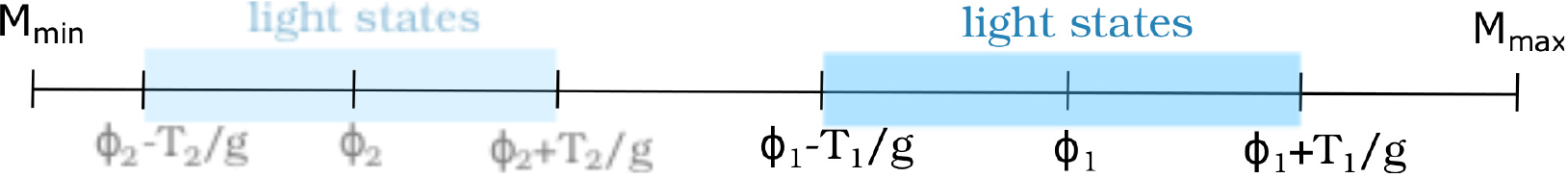}
\caption{Schematic representation of the time evolution in the model assuming that the temperature decreases.}
\label{model-picture}
\end{figure}

In general the mass of states labeled by $i$ is 
as discussed in Sect.~\ref{Supersymmetric distributed mass model}
$m_{\chi_{i}}^2 = m_{\psi_{\chi_{i}}}^2 = g^2(\phi - M_i)^2$
and so is governed by the parameter $M_i$. It is the distribution
of these mass sites $M_i$ that the underlying theory should
determine, but here we treat them as phenomenological parameters
and examine different types of distributions of $M_i$ over mass
sites labelled by $i$.  To understand the meaning of such
mass sites, suppose the inflaton field $\phi$ sat in
the middle of some mass sites.  Thus for all mass sites
$\phi - T/g < M_i < \phi + T/g$, the corresponding fields
$\chi_{i}$ and $\psi_{\chi_{i}}$ would be thermally excited.
As some terminology, we will call the region $\Delta \phi$ surrounding
the inflaton $\phi$ that has thermally excited fields, a thermal
interval. Some examples of this are shown in Fig.~\ref{model-picture}.  
The idea of warm inflation in such models is that
the inflaton rolls through a region with many such mass sites,
thus thermally exciting for some time a given field
and then once $\phi$ is far enough away, that field again is no
longer thermally excited. This implies
as the inflaton slow-rolls during inflation, a thermal interval
surrounding it moves with it.

With no underlying theory to dictate such mass distributions, we
will simply consider one type of construction to make the idea more
tangible.  We will assume the mass sites can be written as
\begin{equation}
M_i = \phi + i \Delta M(\phi,T,m,g) \;,
\end{equation}
where $\Delta M$ is the mass gap in the tower, which may in general be a dynamical function of the inflaton's expectation value and the ambient temperature, as well as of some intrinsic mass scale $m$ and couplings such as $g$. For instance, in a theory with compact extra-dimensions, as is the case of string/M-theory, the mass gap in the Kaluza-Klein tower is inversely proportional to the size of the extra-dimensions, generically set by the expectation value of one or more moduli fields. If such moduli interact with the inflaton and/or the thermal bath, their expectation value may become dynamical (although slowly evolving) during inflation, exhibiting a $\phi$- and/or $T$-dependence\footnote{Note that the bare masses of the fields in the tower relevant for the dissipative process, $gM_i\sim g\phi$, are typically very large and that, at any given time, only a finite number of states become light due to the inflaton field reducing their mass. Therefore, this scenario does not require a low KK/string scale to be implemented in the context of string/M-theory.}. With this motivation in mind, we will take an effective field theory approach and consider functional forms of the mass gap that lead to simple forms for the dissipative coefficient, analyzing their dynamical and observational consequences.

In the expression above $i$ is an integer ranging from mass sites closest
to $\phi$ at $i=1$ up to a maximum mass site that is thermally
excited determined by $i_{max} \Delta M = T/g$.  During the course
of the observable range of inflation (around 50-60 e-folds), 
the inflaton will traverse some distance covering
many thermal intervals $\Delta \phi$
and in general one needs mass distributions where
$\Delta M \ll \Delta \phi$, so that the inflaton crosses many mass sites
in each thermal interval.  
One expects at least tens of mass sites
per thermal interval and hundreds to thousands of thermal
intervals crossed during the observable period of inflation.
In other words, in each e-fold of inflation one expects tens or
few hundred mass sites to be crossed.  Thus thousands
of string states will be thermally excited for
a brief period of time in the course of such a warm inflation period.

With this form of mass distribution function, one can then obtain expressions
for the dissipative coefficient and the effective potential.
For the dissipative coefficient the sum that is required is
\begin{equation}
\sum_{i=1}^{i_{max}} (\phi - M_i)^2 = \sum_{i=1}^{i_{max}} i^2 \Delta M^2 \approx i_{max}^3 \Delta M^2 \simeq T^{3}/(g^{3} \Delta M) \;.
\end{equation}
For example if we wanted 
$\Upsilon \sim \phi$ it means considering $\Delta M \sim T/(g^3 \phi)$
and similarly for other forms of the dissipation function.

In the landscape context, the idea being there are a huge number
of different vacua and some subset can produce a distributed mass
model and some some subset of those could have a distribution of
mass sites that gives a dissipative coefficient that leads
to observably consistent warm inflation.
Since the possible vacua is so huge, its simply seen as a statistical
possibility that somewhere in this landscape a mass distribution of
a particular type can be found.  One would need to see this in
an anthropic way that there could be inflation of many different
forms occurring over the landscape, but some will occur in a way
favorable to create a Universe like ours.  In this respect
choosing a function $\Delta M$  boils down to finding the type
of functions that can lead to a Universe that looks like ours.
If that could successfully be realized, as we will
examine in the next Section, then this model
would have phenomenological viability.  One could then
explore as a much bigger step if its possible to find such
particular types of DM models from string theory, although
we will not be exploring that question here.

We can follow the above prescription in order to evaluate the effective 
finite temperature potential, where for the sake of simplicity we evaluate only the bare masses of the $\chi$'s and $\psi_{\chi}$'s fields. Hence both bosonic and fermionic sector, can be estimated using \cite{Dolan:1973qd,Kapusta:2006,Cline:1996mga} (see eqs.~(\ref{VT-scalars-C1}) and (\ref{VT-fermions-C1}))):
\begin{eqnarray}
V_{T}^{\chi_{i}} &\simeq& \frac{T^{5}}{g\Delta M}\left\{-\frac{\pi^{2}}{45}+\frac{1}{12}+\frac{1}{6\pi}-\frac{1}{32\pi^{2}}\left[\ln\left(\frac{\mu^{2}}{T^{2}}\right)-c_{b}\right] \right\} \,,\label{VT-scalar-discrete} \\
V_{T}^{\psi_{\chi_{i}}} &\simeq&  \frac{T^{5}}{g\Delta M}\left\{-\frac{7\pi^{2}}{360}+\frac{1}{24}+\frac{1}{32\pi^{2}}\left[\ln\left(\frac{\mu^{2}}{T^{2}}\right)-c_{f}\right] \right\}  \label{VT-fermion-discrete} \,,
\end{eqnarray}  
hence if we select $\Delta M \sim T/(g^3 \phi)$ the total effective finite temperature potential becomes:
\begin{equation}\label{total-effective-potential-discrete}
V_{T}^{\chi_{i}}+V_{T}^{\psi_{\chi_{i}}} \simeq  \frac{g^{2}T^{3}}{24}\phi \left[-\pi^{2}+3+\frac{4}{\pi}+\frac{3}{4\pi^{2}}(c_{b}-c_{f})\right] \,.
\end{equation}
Since the thermal corrections to the inflaton potential are, for this particular case, linear in the field, they contribute to the $\epsilon_{eff}$ parameter and need to be taken into account. However, in the scenarios with a quartic chaotic potential $V(\phi)=\lambda\phi^4$ that we analyze in more detail below, we typically find $T/\phi \sim 10^{-5}$ and $\lambda\sim 10^{-15}-10^{-14}$, such that these thermal corrections give a negligible contribution to both the effective potential and its first derivative for any value of the coupling $g$ in the perturbative regime. In addition, in the scenarios where the mass gap is field-independent, thermal corrections to the inflaton potential do not modify the slow-roll parameters.


\section{Results}\label{Results}

Let us now analyze the inflationary dynamics for a quartic scalar 
potential, $V(\phi)=\lambda\phi^4$, taking into account both scalar 
and fermionic contributions to the dissipation coefficient. 
In addition different types of mass distribution functions will
be used leading to dissipative coefficients of various forms, namely
 $\Upsilon\propto T^{2}$, $\Upsilon\propto T$ and $\Upsilon\propto\phi$.  
Finite temperature corrections to the effective potential are computed 
for each case. These effects can be controlled
thus preventing thermal effects 
from generating large contributions to the inflaton mass that 
could reintroduce the ``$\eta$-problem".  

\subsection{$\Upsilon\propto T^{2}$}\label{Evenly distributed mass sites}

For a homogeneous distribution of states in the tower, with constant mass splitting $\Delta M$, as proposed by \cite{Berera:1998px,Berera:1999px,Berera:1999ws}, the resulting scalar coefficients grows with temperature as $\Upsilon\propto T^{2}$, with fermions contributing only about 20$\%$ to the total dissipation as first estimated in \cite{Berera:1998px,Berera:1999px,Berera:1999ws}. Due to the adiabatic condition $\bar{\Gamma}_{\chi}^{S}/H>1$, one requires large values of $T_{*}/H_{*}$; quantitatively we have that at least 
$T_{*}/H_{*}\gtrsim 150$, which in turn yields values of  $Q_{*}\gtrsim 10$. Our results here are consistent with those reported in 
\cite{Berera:1998px,Berera:1999px,Berera:1999ws}. However in the strong dissipation regime it is now understood that the primordial perturbations have a growing mode \cite{Moss:2007cv,Graham:2009bf, Ramos:2013nsa, Bastero-Gil:2014jsa}.  The consequences of this growing mode are to tilt the spectrum of perturbations towards the blue-tilted region, $n_s>1$, and so inconsistent with the CMB data.  Thus, the DM model with a homogeneous mass distribution
resides outside the observational window provided by Planck data \cite{Planck}. Although this particular mass distribution is not consistent with observation, the basic idea of the model shows appealing features and one can explore other types of mass distributions as we will now do. 


\subsection{$\Upsilon=\tilde{C}_{\phi}\phi$ and $V(\phi)=\lambda\phi^4$}

We want to study this model within the strong dissipative regime, since we expect that this model fits perfectly in it. This facilitates all calculations, so we will use standard analytical tools, i.e. we may implement the standard slow-roll parameters $\epsilon_{\phi}=M_{P}^{2}(V_{,\phi}/V)^{2}/2$ and $\eta_{\phi}=M_{P}^{2}V_{,\phi\phi}/V$. Also we introduce another slow-roll parameter to take into account the variation of $\Upsilon$:
\begin{equation}\label{beta-slow-roll}
\beta=M_{P}^{2}\frac{\Upsilon_{,\phi}V_{,\phi}}{\Upsilon V} \,,
\end{equation}
and the slow-roll conditions are now given by $\epsilon_{\phi}<Q\,,|\eta_{\phi}|<Q\,,\beta_{\phi}<Q$. We use the slow-roll equations, at $Q\gg 1$, in order to find a direct relation between $Q$ and $\phi$, yielding:
\begin{equation}\label{Q-Upsilon-phi}
Q=\frac{\tilde{C}_{\phi}}{\sqrt{3\lambda}}\frac{M_{P}}{\phi}\,,
\end{equation}
and the slow-roll parameters are:
\begin{equation}\label{slow-roll-chaotic-Upsilon-constant}
\epsilon_{\phi}=\frac{8M_{P}^{2}}{\phi^{2}} \,,\quad \eta_{\phi}=\frac{12M_{P}^{2}}{\phi^{2}} \,, \quad \beta_{\phi}=\frac{4M_{P}^{2}}{\phi^{2}}
\end{equation}
so in the strong dissipative regime inflation ends at $\epsilon_{\phi}\simeq Q$, this yields the inflaton value:
\begin{equation}
\phi_{end}=\frac{8\sqrt{3\lambda}}{\tilde{C}_{\phi}}M_{P} \,.
\end{equation}
The number of e-folds during inflation in the aforementioned regime is:
\begin{eqnarray}
N_{e}&=&-\frac{1}{M_{P}^{2}}\int_{\phi_{*}}^{\phi_{end}}d\phi\frac{QV}{V_{\phi}}=-\frac{\tilde{C}_{\phi}}{4\sqrt{3\lambda}M_{P}}\int_{\phi_{*}}^{\phi_{end}}d\phi \nonumber\\
&=& \frac{\tilde{C}_{\phi}}{4\sqrt{3\lambda}}\left[\left(\frac{\phi_{*}}{M_{P}}\right)-\left(\frac{\phi_{end}}{M_{P}}\right)\right] \,,
\end{eqnarray}
hence the inflaton value at horizon crossing is:
\begin{equation}
\phi_{*}=\frac{4\sqrt{3\lambda}(N_{e}+2)}{\tilde{C}_{\phi}}M_{P} = 2 \sqrt{\frac{N_e +2}{Q_*}} M_P \,. 
\end{equation}
Once we have determined $\phi_{*}$ we can evaluate the observables at 50-60 e-folds before inflation ends. Before we proceed, one should note that in the strong dissipative regime many parameters and formulas simplify, in fact, one of them is an analytic approximation to the density perturbation amplitude. Recall that for $T>H$, the dominant contribution to the primordial perturbation spectrum are thermal fluctuations of the inflaton field, as opposed to the conventional quantum fluctuations in cold inflation models. Upon exiting the horizon these thermal fluctuations freeze out as classical perturbations and during slow-roll at $Q\gg1$ the amplitude of the curvature perturbation power spectrum is given by \cite{Hall:2003zp,Bartrum:2013oka}:
\begin{equation}\label{power-spectrum-Qbb1}
\Delta_{\mathcal{R}}^{2}\simeq \frac{9}{4\pi^{2}}\frac{H_{*}^{5}T_{*}Q_{*}^{5/2}}{V_{\phi_{*}}^{2}} \,,
\end{equation}
where all quantities are evaluated at horizon-crossing. Hence the scalar spectral index $n_{s}-1\simeq d\ln\Delta_{\mathcal{R}}^{2}/dN_{e}$ is modified as well, becoming \cite{Hall:2003zp}: 
\begin{eqnarray}
n_{s}&=&1+\frac{3}{2}\left(\frac{\eta_{V}}{Q}-\frac{3}{2}\frac{\epsilon_{V}}{Q}-\frac{3}{2}\frac{\beta_{V}}{Q}\right)=1-\frac{9}{4}\left(\frac{2\sqrt{3\lambda}}{\tilde{C}_{\phi}}\frac{M_{P}}{\phi_{*}}\right)  \nonumber\\
&=& 1-\frac{9}{4}\left(\frac{1}{N_{e}+2}\right) \,.
\end{eqnarray}
Remarkably, note that $n_{s}$ only depend of the number of e-folds: $n_{s}(N_{e}=50)=0.9567$ and $n_{s}(N_{e}=60)=0.9637$; where at 60 e-folds this scalar spectral index agrees outstandingly with Planck data \cite{Planck}. The tensor-to-scalar ratio $r=\Delta_{t}^{2}/\Delta_{\mathcal{R}}^{2}$, as mentioned above, is typically reduced by the modifications to the scalar curvature perturbations introduced due to dissipation, which is basically a function of $Q_{*}$. We illustrate this fact by using the slow-roll dynamics, where the ratio $r$ can be written as:
\begin{equation}\label{tenso-to-scalar-ratio-Upsilon-phi}
  r=\frac{2H_{*}^{2}}{\pi^{2}M_{P}^{2}\Delta_{\mathcal{R}}^{2}}=\frac{2\lambda}{3\pi^{2}\Delta_{\mathcal{R}}^{2}}\frac{\phi_{*}^{4}}{M_{P}^{4}}=\frac{32\lambda}{3\pi^{2}\Delta_{\mathcal{R}}^{2}}\frac{(N_{e}+2)^{2}}{Q^{2}}
%
\end{equation}
%
The value of $\lambda$ is fixed by using the normalization of amplitude of the primordial spectrum $\Delta_{R}^{2}\simeq 2.2\times 10^{-9}$ \cite{Planck}. Using the slow-roll equations in Eq. (\ref{power-spectrum-Qbb1}), we have:
\be
\lambda = (\Delta_{\mathcal{R}}^{2})^{4/3} \left(\frac{(12\pi^{4}C_{R}^{1/2} )^{2/3}}{(N_e+2)^3}\right) \,,
\ee
where $C_R= \pi^2 g_{eff}/30$, and $g_{eff}= 1 + 15 N_M/4$, $N_M$ being the no of bosonic $\chi_i$ (fermionic $\psi_i$) light degrees of freedom at horizon crossing. The tensor-to-scalar ratio is then just a function of $Q_*$ and $N_e$,
\be
r=\frac{64 ( 6\Delta_{\mathcal{R}}^{2} )^{1/3}}{3 Q_*^2 (N_e+2)}  \,.\ee 
As one expected, the tensor-to-scalar ratio is highly suppressed by dissipation. In fact this ratio lies in the region $10^{-9}< r\lesssim 10^{-4}$ at $10\leq Q_{*}\leq 1000$, for $N_M=O(10-100)$, having the bigger suppression at the largest $Q_{*}$.     

\begin{figure}[h!] 
\includegraphics[scale=1.0]{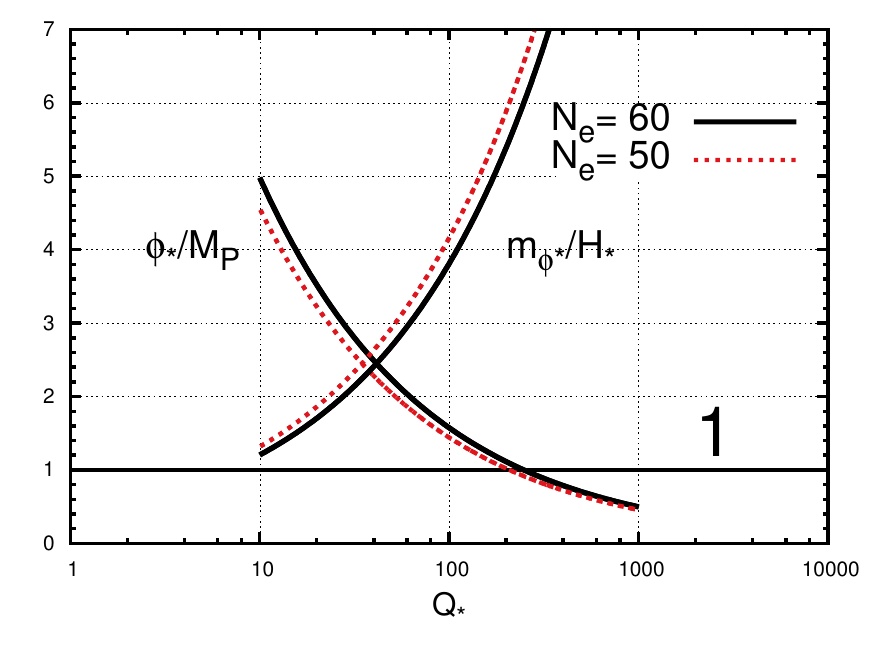}
\caption{Ratios $m_{\phi_{*}}/H_{*}$ and $\phi_{*}/M_{P}$, as a function of the dissipative ratio $Q_{*}$, within a DM scenario with a dissipative coefficient of the form $\Upsilon\propto\phi$ described by a quartic potential for 50 (red dashed-line) and 60 (black solid-line) e-folds of inflation. Note that $m_{\phi_{*}}/H_{*}$ is always larger than 1 for $Q_* > 10$, while $\phi_{*}/M_{P}<1$ (sub-Planckian) at $Q_{*}\sim 210$.}\label{fig:phi_mphi_Qstar}
\end{figure}

\begin{figure}[htbp] 
\includegraphics[scale=0.4]{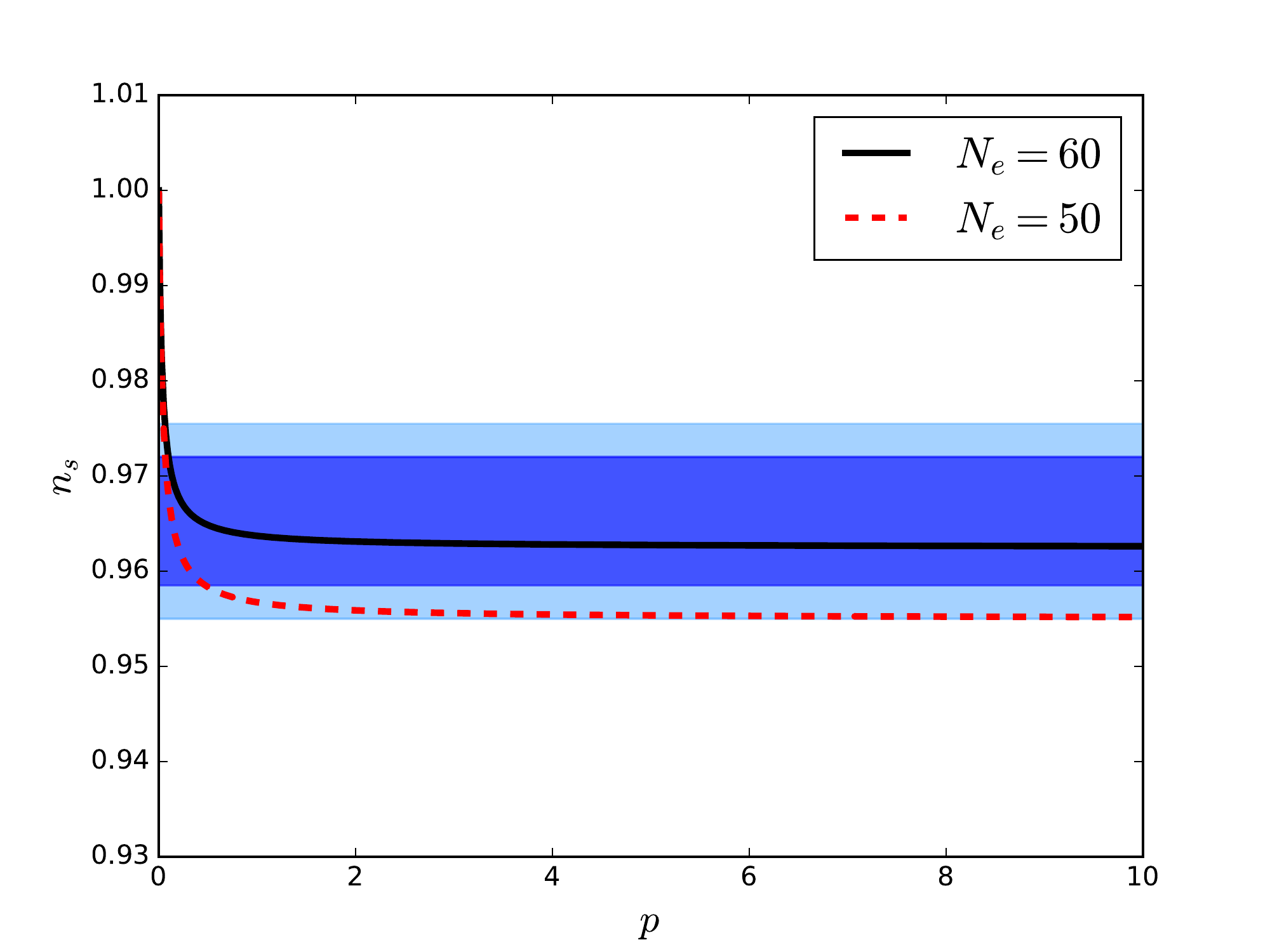}  
\includegraphics[scale=0.4]{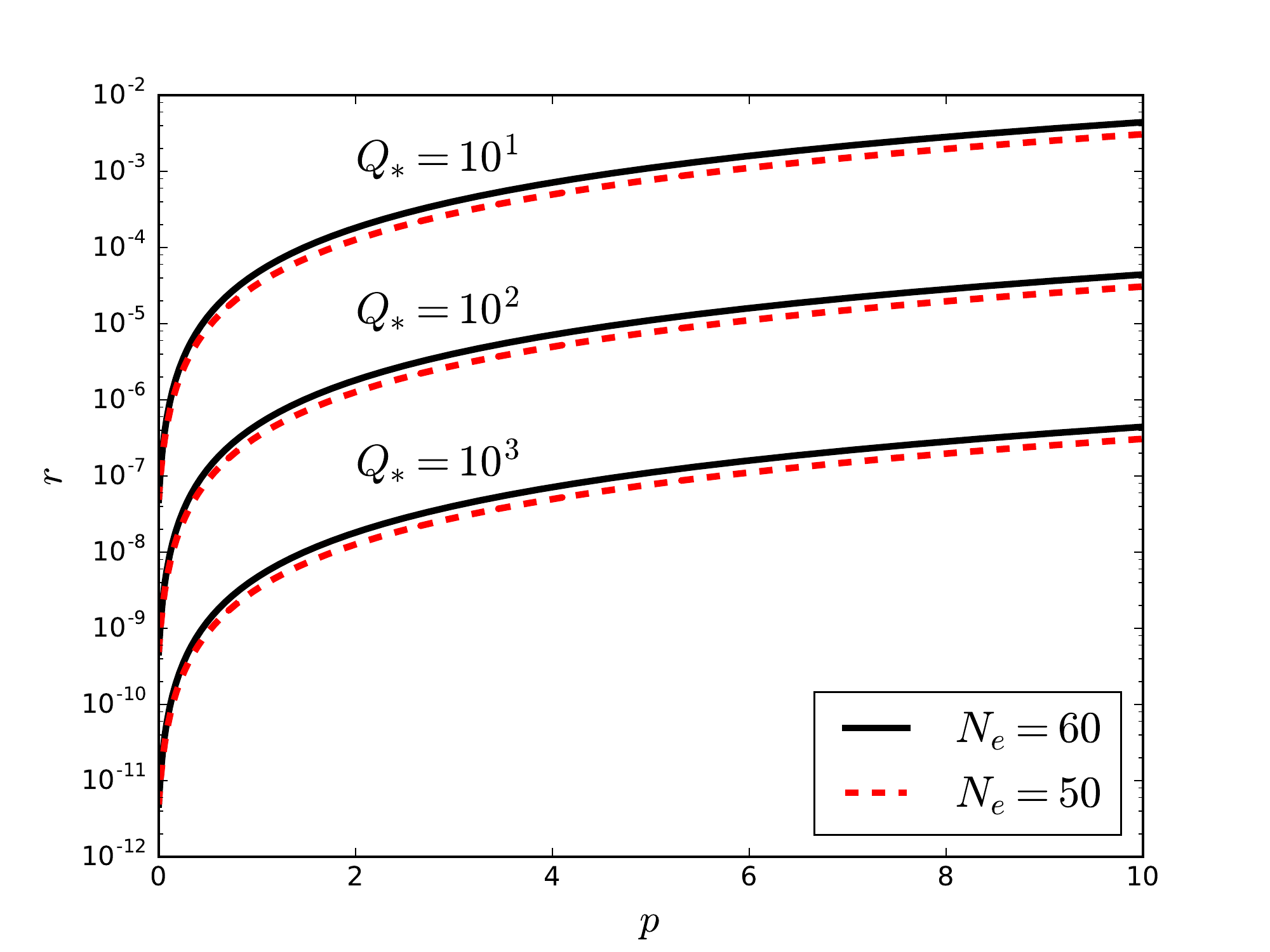} 
\caption{Observational predictions of the DM scenario with a dissipative coefficient of the form $\Upsilon\propto\phi^{p}$ described by a quartic potential for 50 (red dashed-line) and 60 (black solid-line) e-folds of inflation. The plot on the left shows the spectral index $n_{s}$ as a function the parameter $p$, while the plot on the right shows the tensor-to-scalar ratio $r$ as a function of $p$ for three values of $Q_{*}$. The blue contours correspond to the $68\%$ and $95\%$ C.L. results from Planck 2018 TT,TE,EE+lowE+lensing data \cite{Planck}. Note that $n_{s}$ lies well inside the Planck contours at $p\geq 1$. On the other hand, $r$ gets enhanced for larger $p$. }\label{fig:quartic_general_ns-Qstar-r}
\end{figure}

We can also look at the ratios $\phi_{*}/M_{P}$ and $m_{\phi_{*}}/H_{*}$, in order to illustrate that for this model the relevant scales can indeed happen in the  sub-Planckian region, as shown in Fig.~\ref{fig:phi_mphi_Qstar}. This occurs thanks to the high dissipation dynamics, since in WI field potentials that are not flat enough to allow the standard slow-roll inflaton evolution, i.e. they do not have large enough initial inflaton values so not ``sufficient" inflation takes place, can in fact lead to a longer period of inflation due to the extra friction induced by $\Upsilon$. Hence, for small (even sub-Planckian) $\phi_{*}$ inflation can last 50-60 e-folds in the strong dissipative regime.
We can even generalise above results for a dissipative coefficient of the form: $\Upsilon=C_{\phi}\phi^{p}m^{1-p}$, where $m$ is a mass scale and $p$ is a free parameter. For this generic case, the scalar spectral index and the tensor-to-scalar ratio become very simple expressions that yield a remarkable agreement with Planck legacy \cite{Planck} (see Fig.~\ref{fig:quartic_general_ns-Qstar-r}): 
\begin{equation}\label{ns-Upsilonphi-generic}
n_{s}= 1-\frac{9}{4}\left(\frac{p}{pN_{e}+2}\right) \,,\qquad r=\frac{32\lambda}{3\pi^{2}\Delta_{\mathcal{R}}^{2}}\frac{(pN_{e}+2)^{2}}{Q^{2}}
\end{equation}

The good agreement between WI with a dissipative coefficient of the 
form: $\Upsilon=C_{\phi}\phi^{p}m^{1-p}$ described by a chaotic quartic 
potential has been reported before, for instance \cite{Ramos:2013nsa} 
analyzed the $\Upsilon\propto\phi^{2}$ case, finding observably favorable region
for $Q\gg1$. On the other hand, the tensor-to-scalar 
ratio is enhanced by larger $p$, so for very 
large $p$ this ratio $r$ can indeed become bigger than 1.
The limits on ongoing 
and planned B-mode polarization experiments \cite{Kamionkowski:2015yta} 
suggest sensitivity down to $r\gtrsim 10^{-3}$.  This
leaves open the door to access desirable 
values of $r$ for very large $Q_{*}$.


\subsection{$\Upsilon=C_{T}T$ and $V(\phi)=\lambda\phi^4$}

We can ensure a linear dissipative coefficient by considering a DM model 
with a constant no. of bosonic (fermionic) light degrees of freedom 
$N_M$ during the evolution of the inflaton field. 
This is equivalent at having in the continuum limit a density 
of states that goes inversely with $T$, such  that 
$N_M \simeq   2 T n(\phi) /g$. Using this, and perfoming 
the sums in Eqs (\ref{Upsilon_DMS}) and (\ref{Upsilon_DMF}) 
in the continuum limit, we then have:
\bea
\Upsilon^{S}(\phi,T) &=& N_M C_{T}^{S}T\,, \quad C_{T}^{S}=\frac{2g^{2}\sqrt{f}}{\pi}\ln\left(\frac{2}{\sqrt{f}}\right)\left[1-\frac{f}{4}\text{ArcTan}\left(\frac{4}{f}\right)\right] \,, \\
\Upsilon^{F}(\phi,T) &=& N_M C_{T}^{F}T \,, 
\end{eqnarray}
where $C_T^{F}$ was given in Eq. (\ref{Upsilon_DMF}). The observational predictions will depend now on the couplings $g$ and $h$, and the parameter $N_M$. 

To ensure the consistency of the analysis, namely the computation of 
the dissipation coefficients, we must verify that both the bosonic 
and fermionic fields coupled to the inflaton are kept close to thermal 
equilibrium, thus requiring 
$\bar{\Gamma}_{\chi_i}, \bar{\Gamma}_{\psi_{\chi_i}}> H$. Since both 
average decay widths are proportional to the temperature and smaller than 
the latter for perturbative couplings, if the latter conditions are
satisfied we also ensure that $T\gtrsim H$ throughout inflation, thus 
validating the flat space approximation used in computing the dissipative 
coefficients. For the quartic potential, the ratio $T/H$ increases during
inflation, being proportional to the dissipative ratio $Q$, such that we 
only have to ensure that these conditions are met at the moment of 
horizon-crossing for the relevant CMB scales.

We note that in another realization of warm inflation with light fields, the Warm Little Inflaton (WLI) scenario, one has to impose additional constraints on the temperature during warm inflation so as to keep it above the mass of the fermionic fields coupled to the inflaton and below the underlying symmetry breaking scale. Such conditions are absent in the DM model, which is therefore less constraining. This comes of course at the expense of considering a whole tower of fields coupled to the inflaton, although only a small number of fields contributes effectively to dissipation at any given time.

In our analysis, we have also included the one loop thermal effective potential, computed in Appendix \ref{appendix c}. The bosonic contribution is given in Eq. (\ref{VT-bosons-continuum-uniform}), while the fermionic one is given in Eq. (\ref{VT-fermions-continuum-uniform}). From those expressions, one can see that
the contribution of the finite temperature corrections to the derivatives of the effective potential is absent in this model. We have kept anyway the subdominant thermal contribution to the potential, setting the renormalisation scale such as $\mu=\exp(c_{f}/2)T_*$, where $T_*$ is the temperature at horizon crossing. It is worth mentioning that by choosing the renormalisation scale through another prescription, for instance $\mu=\exp(c_{b}/2)T*$, the outcome is not substantial altered. 

With a $T$-dependent dissipative coefficient, the amplitude of the primordial spectrum Eq. (\ref{general_thermal-spectrum}) includes the growing mode function $G(Q_{*})$. For a chaotic quartic potential, it was obtained numerically in \cite{Bastero-Gil:2016qru, Bastero-Gil:2018uep}:
\begin{equation}\label{general_growing-mode}
G(Q_{*}) \simeq 1+0.0185 \, Q_{*}^{2.315}+0.335 \, Q_{*}^{1.364} \,,
\end{equation}
which alongside the dynamical evolution equations described in the previous section allows us to compute the inflationary observables for the chaotic model for $50-60$ e-folds of inflation, provided that the adiabatic conditions discussed above are satisfied. The inflaton self-coupling $\lambda$ is fixed as before by using the observed amplitude of curvature perturbations, $\Delta_{R}^{2}\simeq 2.2\times 10^{-9}$ \cite{Planck}, at 50-60 e-folds before the end of inflation.
\begin{figure}[h] 
\includegraphics[scale=0.98]{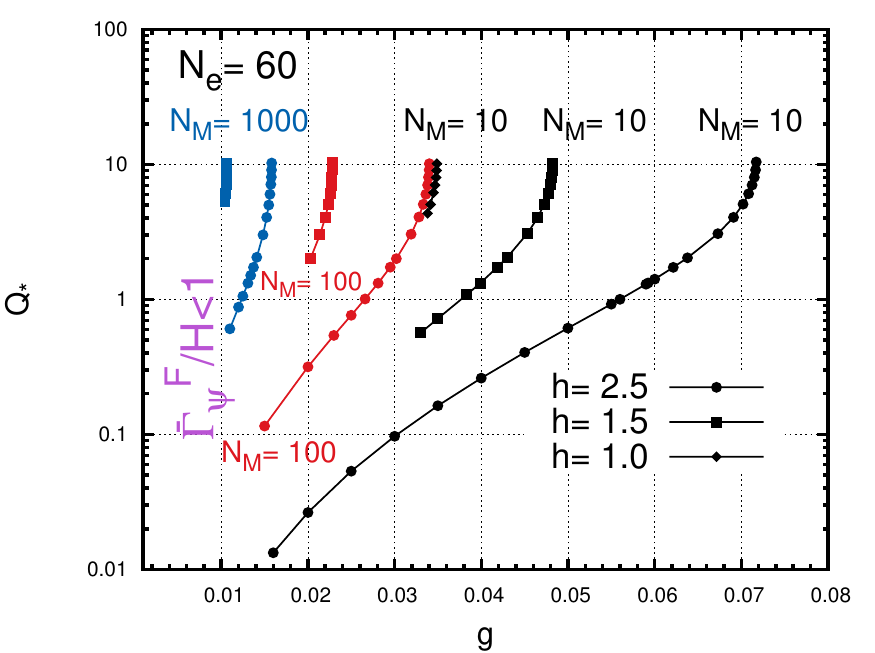}
\includegraphics[scale=0.98]{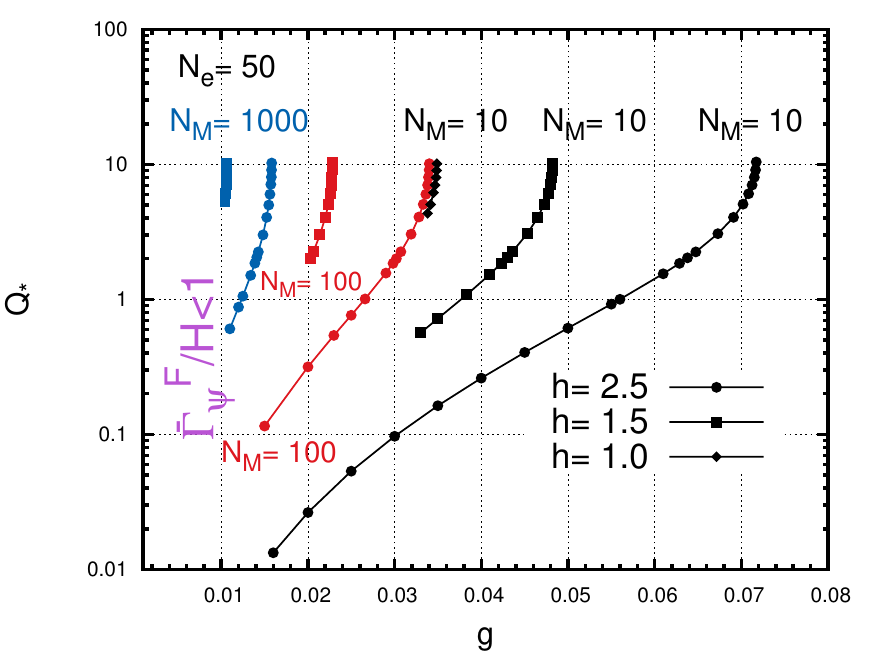}
\caption{Dissipative ratio at horizon-crossing, $Q_*$, as a function of the coupling $g$, within a DM scenario with a dissipative coefficient of the form $\Upsilon\propto T$ described by a quartic potential and $N_e=50$ (right) or $N_e=60$ (left) e-folds of inflation. We consider different numbers of light states $N_M=10, 100$ and 1000, shown by the black, red and blue curves, respectively; as well as different values of the Yukawa coupling $h=2.5$ (circles), $h=1.5$ (rectangles), and $h=1.0$ (diamonds). Note that the adiabatic conditions $\bar{\Gamma}_{\chi_i}/H\gtrsim1$ is not satisfied for small values of $g$.}\label{fig:quartic_g_Q}
\end{figure}

For the quartic potential, we show in Fig.~\ref{fig:quartic_g_Q} the obtained values of the dissipative ratio at horizon-crossing, $Q_*$, as a function of the coupling $g$, for different numbers of light states and values of the Yukawa coupling $h$. We note in this figure that the adiabatic condition $\bar{\Gamma}_{\chi_i}/H\gtrsim1$ for the fermions is only satisfied for $g\gtrsim 0.01$. We also find a lower bound $Q_{*}>0.01$ for the smallest number of fields considered, and that the adiabatic conditions imply generically $h\gtrsim 1$. We only show in this figure values of $Q_*<11$ since, as we will see below, larger values are incompatible with observational data \cite{Planck}.

This behaviour had been reported before for the WLI model \cite{Bastero-Gil:2016qru,Bastero-Gil:2017wwl, Bastero-Gil:2018uep}. For instance, the analysis in \cite{Bastero-Gil:2018uep} presented a rather similar lower bound on $g\gtrsim 0.01$ and $Q*\gtrsim 0.001$ due to the adiabatic condition, while the conditions on the temperature limit this coupling from above, $g\lesssim 0.1$ such that $Q*\lesssim 0.25$. However, the DM model yields a substantially wider consistent parametric range, particularly for the value of the dissipative ratio at horizon-crossing. 

Note that for larger values of the number of light states $N_{M}$ and of the Yukawa coupling $h$ we obtain a much narrower range of consistent values for the coupling $g$, and the lower bound on $Q_*$ also increases with $N_M$ and $h$. We thus find scenarios where inflation can start either in the weak or strong dissipation regimes, noting that $Q$ grows during inflation such that for a wide range of parameters one reaches $Q>1$ before the end of inflation, a necessary condition for radiation to dominate after the slow-roll regime with no further reheating (see eq.~(\ref{radiation_abundance})).

\begin{figure}[h!] 
\includegraphics[scale=0.6]{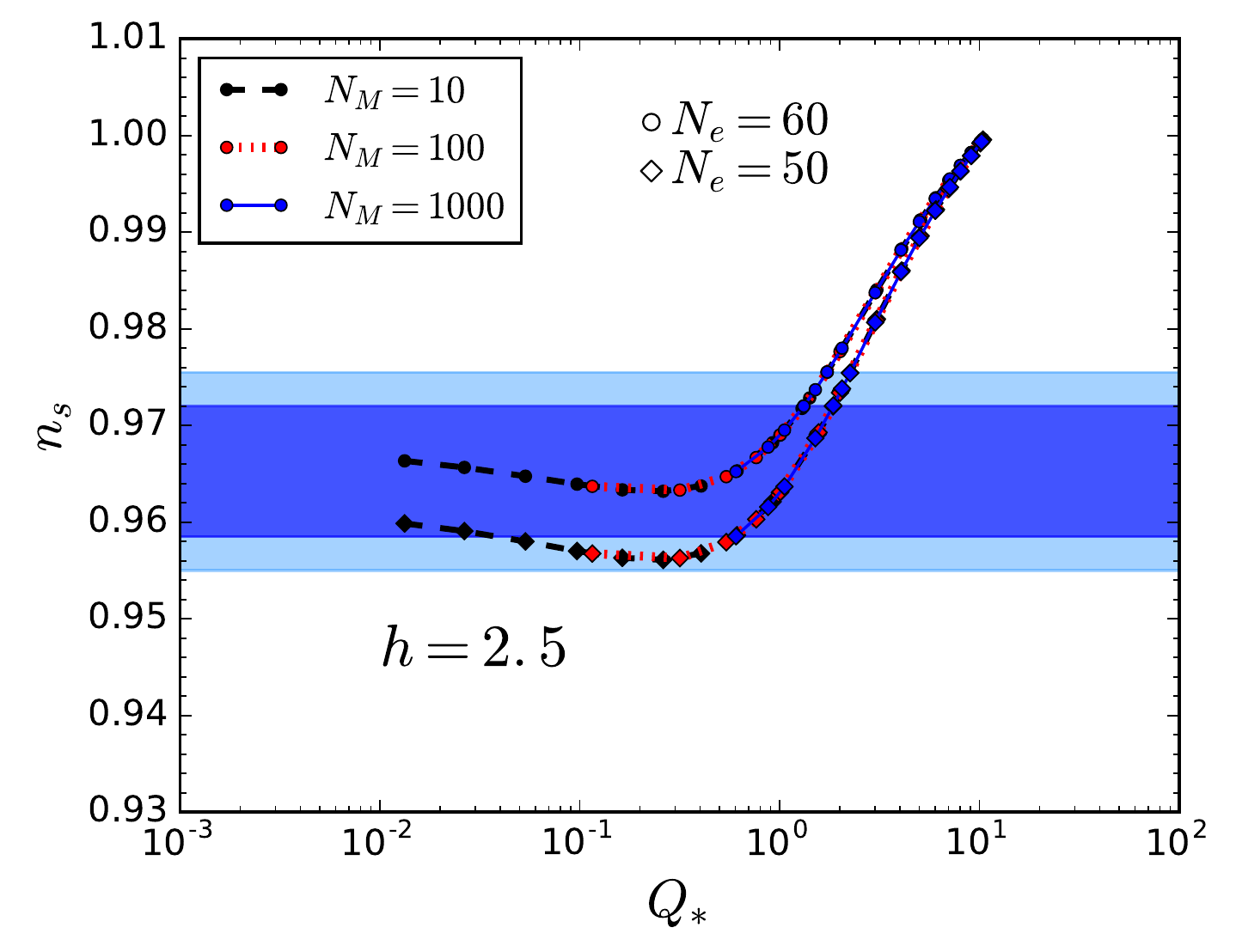}
\includegraphics[scale=0.6]{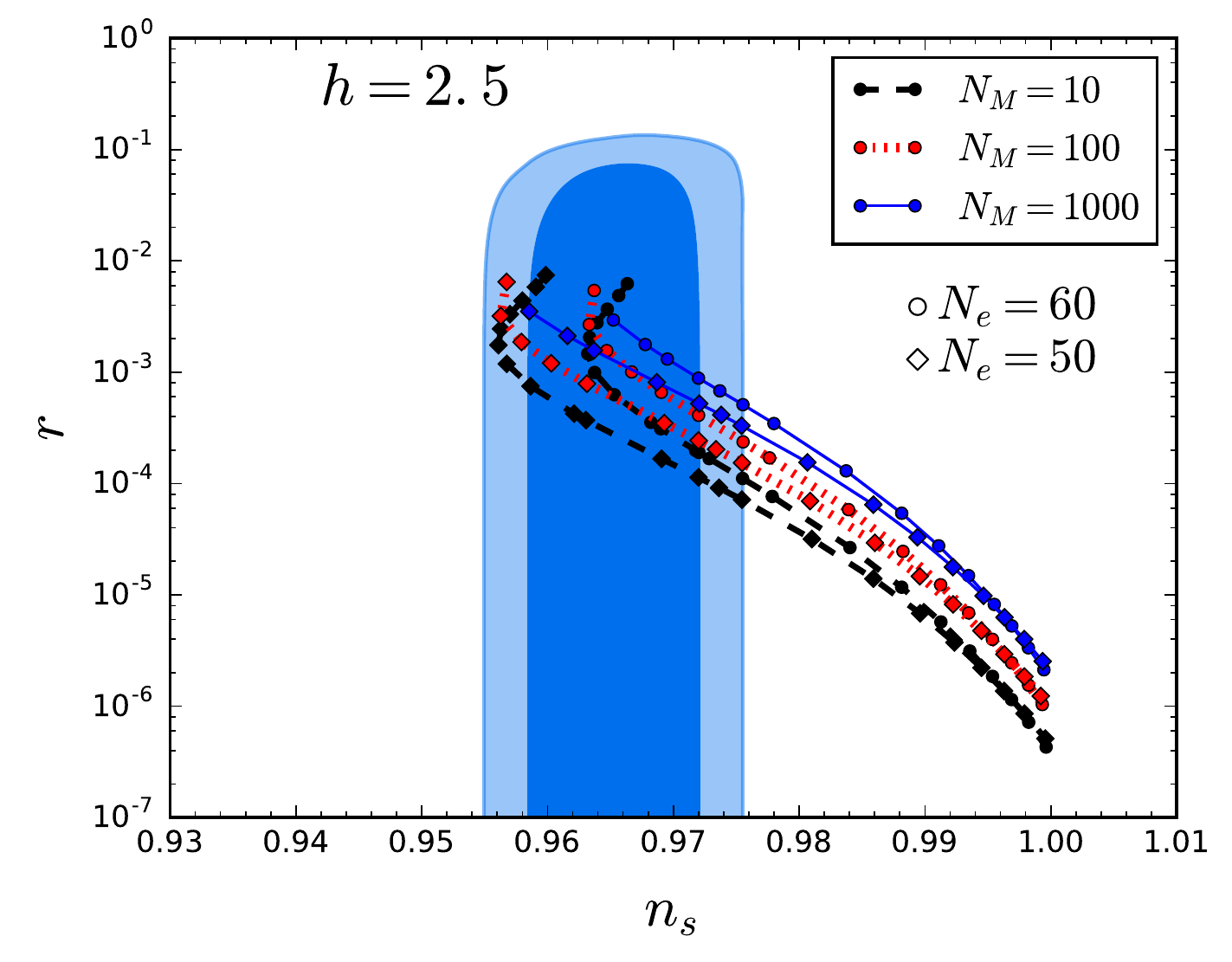}
\caption{Observational predictions of the DM scenario with a dissipative coefficient of the form $\Upsilon\propto T$ described by a quartic potential for 50 (diamonds) and 60 (circles) e-folds of inflation, a Yukawa coupling $h=2.5$, and three different values of the number of light states $N_{M}=10,100,100$ in colours (line-style) black (dashed-line), red (pointed-line), and blue (solid-line) respectively. The plot on the left shows the spectral index $n_{s}$ as a function of the dissipative ratio at horizon-crossing, $Q_*$, while the plot on the right shows the allowed trajectories in the $(n_s,r)$ plane. The blue contours correspond to the $68\%$ and $95\%$ C.L. results from Planck 2018 TT,TE,EE+lowE+lensing data \cite{Planck}.}\label{fig:ns-Qstar-r_h_25}
\end{figure}
\begin{figure}[h] 
\includegraphics[scale=0.6]{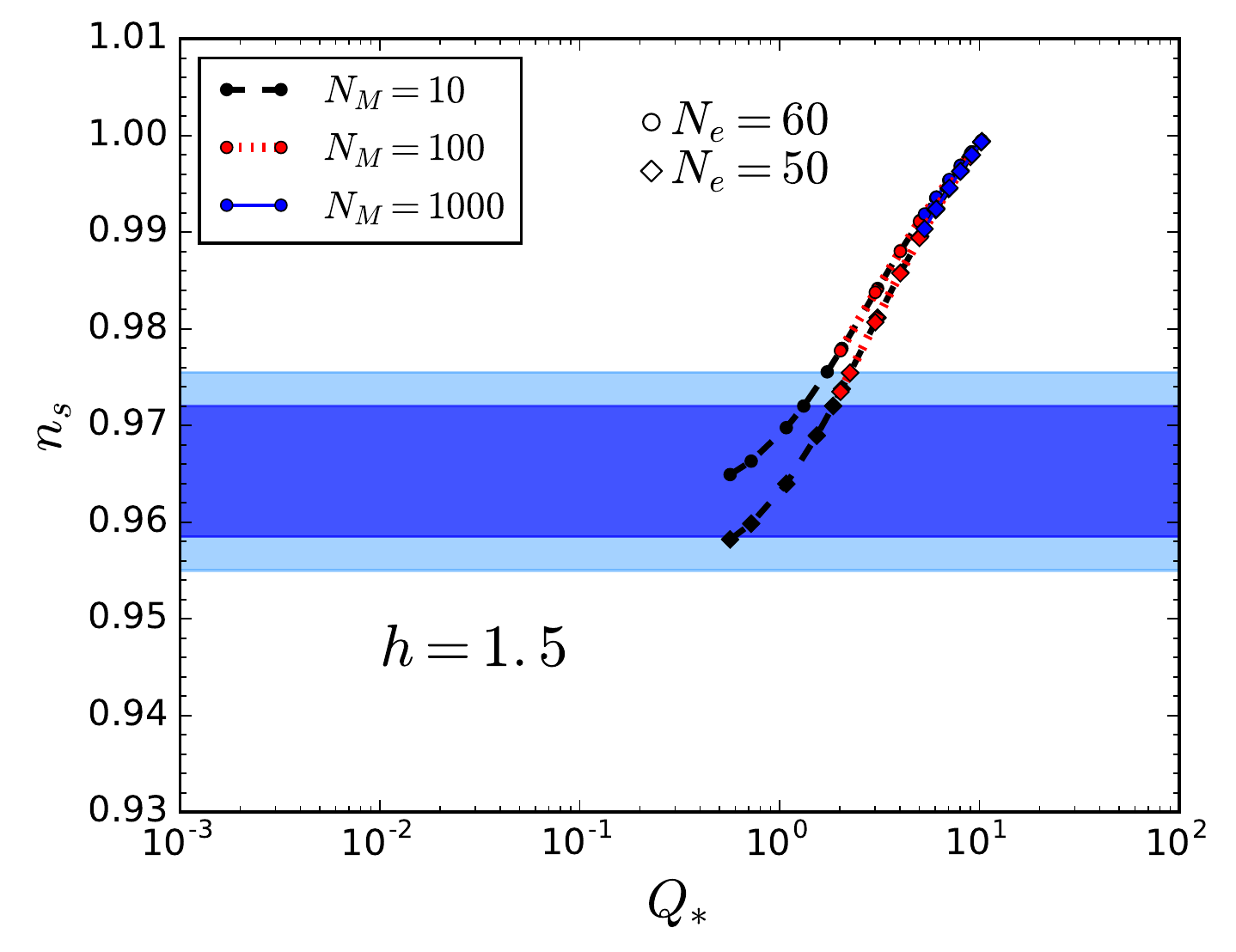}
\includegraphics[scale=0.6]{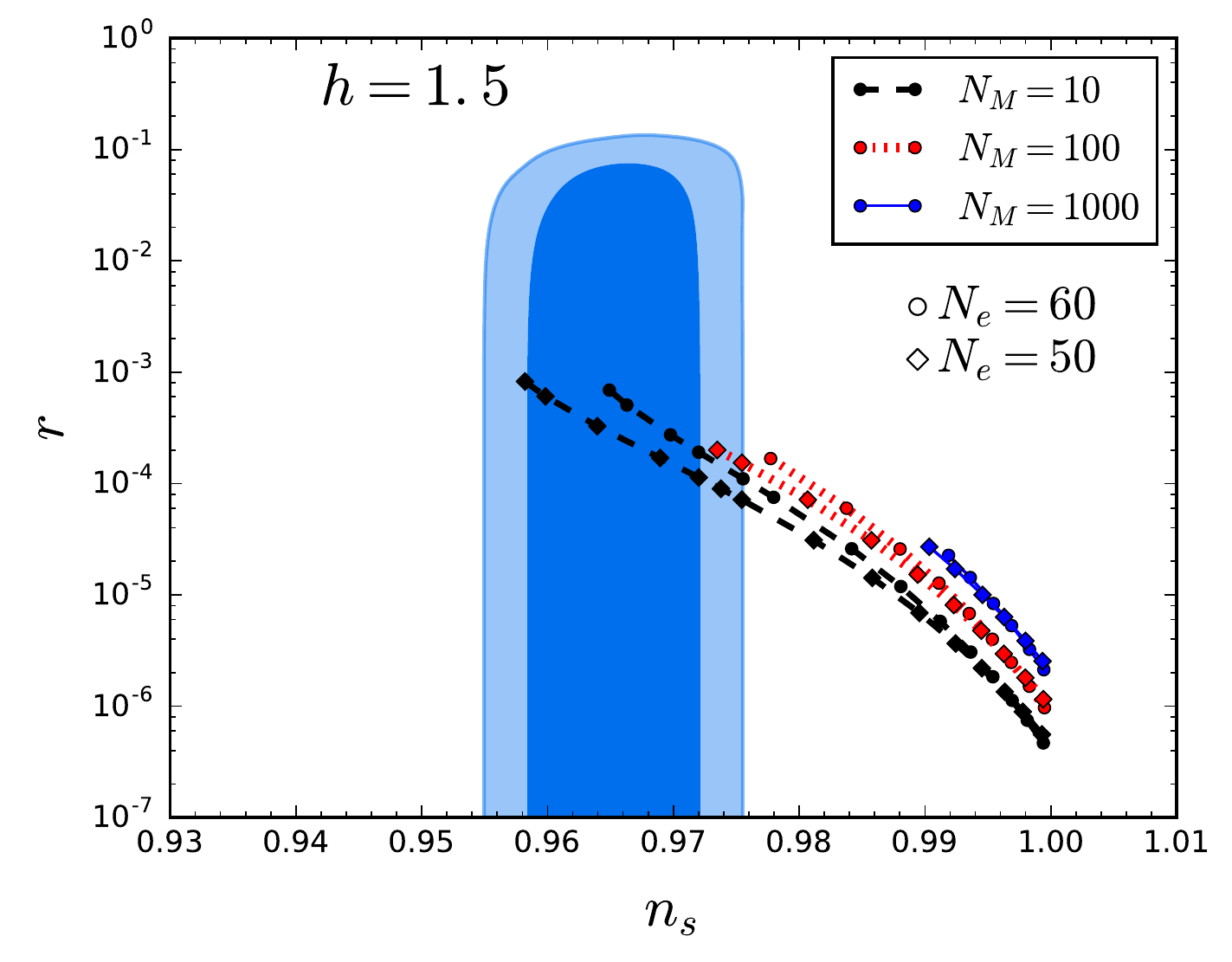}
\caption{Observational predictions of the DM scenario with a dissipative coefficient of the form $\Upsilon\propto T$ described by a quartic potential for 50 (diamonds) and 60 (circles) e-folds of inflation, a Yukawa coupling $h=1.5$, and three different values of the number of light states $N_{M}=10,100,100$ in colours (line-style) black (dashed-line), red (pointed-line), and blue (solid-line) respectively. The plot on the left shows the spectral index $n_{s}$ as a function of the dissipative ratio at horizon-crossing, $Q_*$, while the plot on the right shows the allowed trajectories in the $(n_s,r)$ plane. The blue contours correspond to the $68\%$ and $95\%$ C.L. results from Planck 2018 TT,TE,EE+lowE+lensing data \cite{Planck}.}\label{fig:ns-Qstar-r_h_15}
\end{figure}
\begin{figure}[h!] 
\includegraphics[scale=0.6]{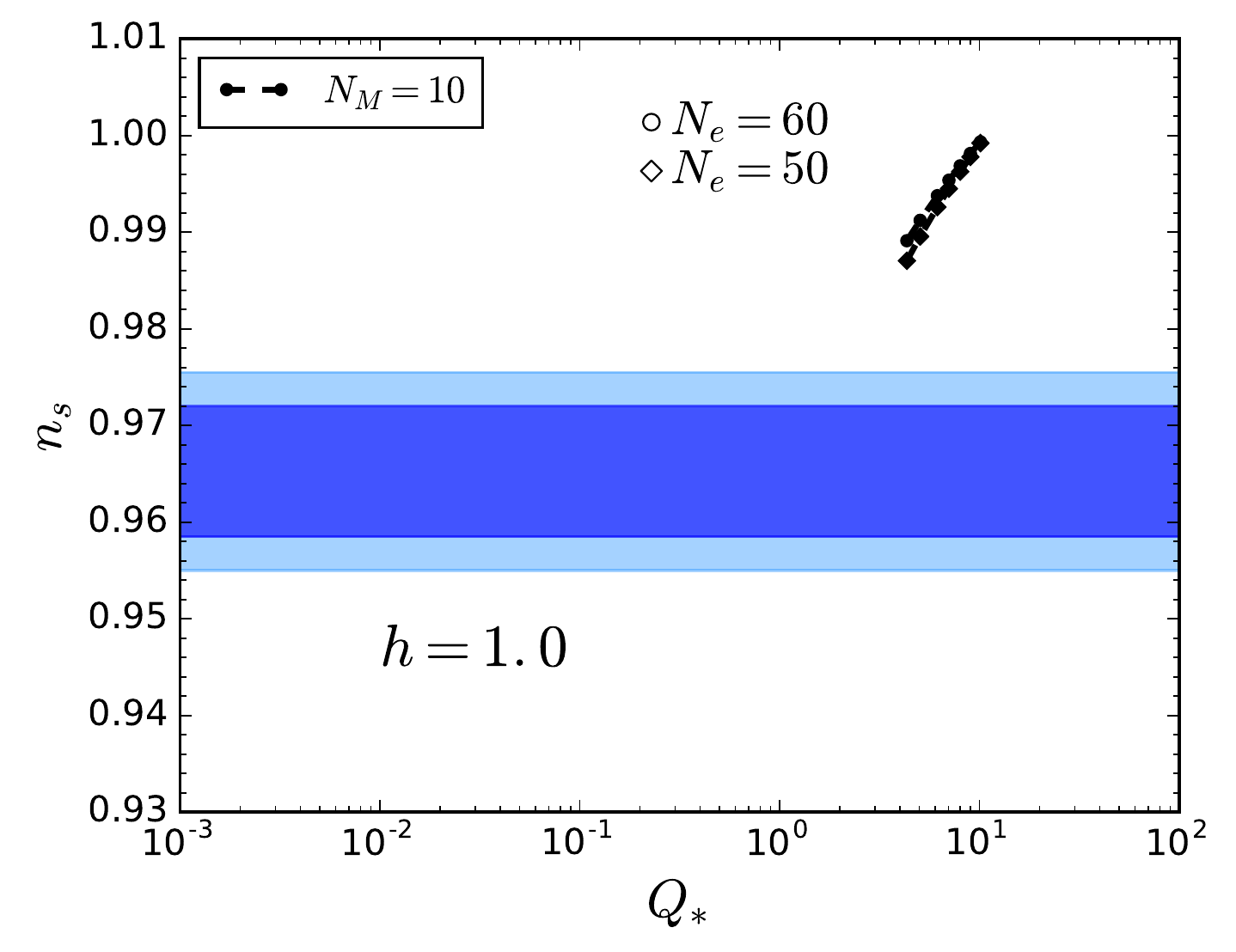}
\includegraphics[scale=0.6]{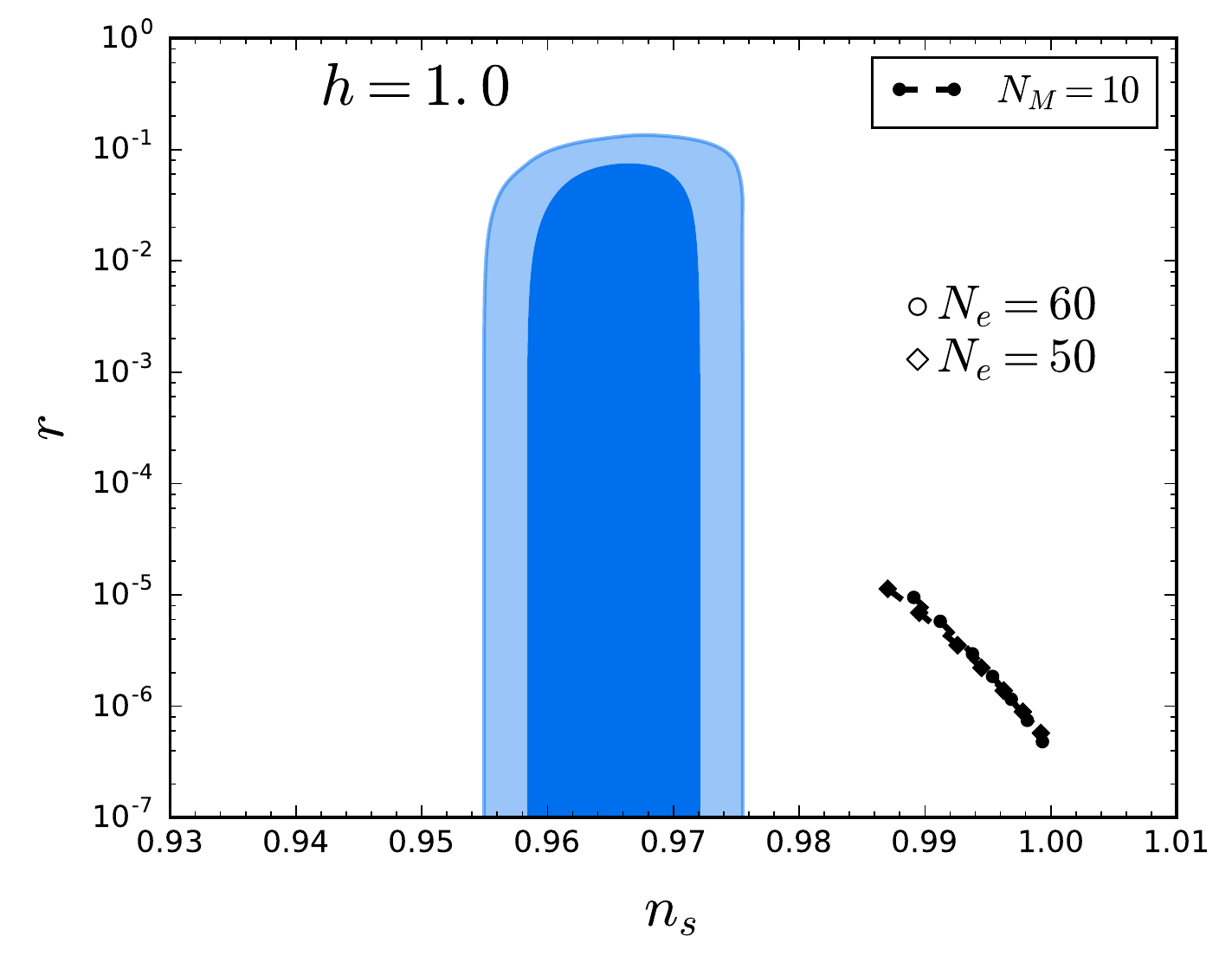}
\caption{Observational predictions of the DM scenario with a dissipative coefficient of the form $\Upsilon\propto T$ described by a quartic potential for 50 (diamonds) and 60 (circles) e-folds of inflation, a Yukawa coupling $h=2.5$, and three different values of the number of light states $N_{M}=10,100,100$ in colours (line-style) black (dashed-line), red (pointed-line), and blue (solid-line) respectively. The plot on the left shows the spectral index $n_{s}$ as a function of the dissipative ratio at horizon-crossing, $Q_*$, while the plot on the right shows the allowed trajectories in the $(n_s,r)$ plane. The blue contours correspond to the $68\%$ and $95\%$ C.L. results from Planck 2018 TT,TE,EE+lowE+lensing data \cite{Planck}.}\label{fig:ns-Qstar-r_h_1}
\end{figure}

In Figs.~\ref{fig:ns-Qstar-r_h_25}, \ref{fig:ns-Qstar-r_h_15} and \ref{fig:ns-Qstar-r_h_1} we show the predictions for the scalar spectral index and tensor-to-scalar ratio in the allowed parametric ranges, for different values of $h$ and $N_M$, exhibiting a remarkable consistency with the Planck legacy results. This is particularly relevant given that the quartic potential is already excluded by such survey within the CI paradigm. 

We find agreement with the Planck legacy data in the parametric ranges yielding values of $Q_*\lesssim 1$, which is easier to achieve for larger values of the Yukawa coupling, as well as a smaller number of light fields. For instance, we cannot find consistency with the Planck data for $h=1$, since smaller values of the Yukawa coupling lead to stronger dissipation at horizon-crossing. These results were expected, since the growing mode in the spectrum makes it more blue-tilted with increasing $Q_*$.

In general, the tensor-to-scalar ratio lies in the range $10^{-4}\lesssim r \lesssim 10^{-2}$, for values of the scalar spectral index within the Planck window, being more suppressed for larger values of $Q_*$ (smaller Yukawa couplings and larger number of light fields $N_M$). This feature is quite generic, since a similar notable trend was obtained in \cite{BasteroGil:2009ec,Cai:2010wt, Bartrum:2013fia,Bastero-Gil:2016qru,Bastero-Gil:2017wwl,Bastero-Gil:2018uep}.  Additionally, this significant agreement with Planck data is easily achieved with only a small number of light states $N_{M}$.

The remarkable agreement between the DM quartic model and the Planck data is similar to what had been obtained for the WLI quartic model \cite{Bastero-Gil:2016qru,Bastero-Gil:2017wwl,Bastero-Gil:2018uep}, where the dissipation coefficient is also proportional to the temperature. However, while the more constraining conditions of the WLI model limit the consistent parametric range to $Q_*\lesssim 0.25$, we have found consistent DM scenarios with up to $Q_*\sim 1.85-2.25\,\,(1.32-1.73)$ for 50 (60) e-folds of inflation within the contours correspond to the $68\%$ and $95\%$ C.L. results from Planck 2018 TT,TE,EE+lowE+lensing data \cite{Planck}. In addition, this noteworthy consistency with Planck, which also leads to a finite range for the tensor-to-scalar ratio, might be relevant in the search of primordial gravitational waves via B-mode polarization experiments in the near future (see e.g.~\cite{Kamionkowski:2015yta}).


Finally we evaluate both swampland criteria within the observational window provided by Planck \cite{Planck}, and we find that none is fully satisfied, with $10\lesssim|\Delta\phi|/M_{P}\lesssim 21$ and $0.19\lesssim M_{P}|V_{eff,\phi}|/V_{eff}\lesssim 0.4$ for 60 e-folds and $7\lesssim|\Delta\phi|/M_{P}\lesssim 19$ and $0.2\lesssim M_{P}|V_{eff,\phi}|/V_{eff}\lesssim 0.55$ for 50 e-folds. Nevertheless, the scenarios for which $Q_*\sim 1$ are closer to satisfying these criteria than the corresponding cold inflation scenarios, and could in fact satisfied them if they were somewhat relaxed. For instance, it has been discussed in \cite{Kehagias:2018uem,Das:2018hqy} that $c$ could be as small as $\mathcal{O}(10^{-11})$, and it does not object perceiving de Sitter vacua in string landscapes. Moreover, there has been some controversy on the precise values for $\Delta$ and $c$, given that they depend on the specific string model being considered \cite{Akrami:2018ylq}. Basically, if $\Delta$ were increased by an order of magnitude, and $c$ were decreasing by around $c\sim0.4$ both swampland criteria could be easily fulfilled for $Q_{*}\gtrsim 1$.


\section{Conclusions}\label{Conclusions}

Warm inflation offers a unique description of the early universe 
cosmology. It is a picture of an inflationary dynamics in which the 
state of the universe during inflation is not the vacuum state, but rather 
an excited statistical thermal state. It introduces dissipation into the 
inflationary dynamics which can be well explained by first principles 
of a quantum multi-field theory. Besides, this approach has several 
attractive features. For instance, the additional friction may ease the 
required flatness of the inflaton potential alleviating the 
swampland criterion. Also, even if radiation is subdominant during 
inflation, it may smoothly become the leading component if $Q\gtrsim 1$ at 
the end of inflation ($\epsilon_{eff}\sim 1+Q$), with no need for a separate 
reheating period\footnote{Although we have not specified how the fields in the tower interact with the Standard Model degrees of freedom, which is a model-dependent issue, these will generically be excited at the end of inflation once the Hubble rate is sufficiently reduced, given the large values of the temperature $T\sim 10^{12-13}$ GeV when radiation becomes dominant. In fact, it would require extremely small couplings between the tower and Standard Model states to prevent the latter's thermalization before nucleosynthesis at $T\sim 10$ MeV (see also \cite{Rosa:2018iff}).}. It also may explain the nature of the classical 
inhomogeneities observed in the CMB, since for WI the fluctuations of the 
inflaton are thermally induced; hence there is no need to explain the 
troublesome quantum-to-classical transition problem of CI, due to the 
purely quantum origin of the CI density perturbations. Furthermore, 
it was shown early on \cite{Berera:1999ws}, 
in fact in studying the DM model, that
the presence of dissipation and temperature lowers the energy scale of
inflation in
monomial models such as $\lambda \phi^4$ in comparison to
the cold inflation results.  This was further developed in
\cite{BasteroGil:2009ec,Cai:2010wt, Bartrum:2013fia}, with the prediction 
for a low tensor-to-scalar ratio, which subsequently has
been shown to be consistent with 
data \cite{Planck}.  These observational predictions for warm inflation were
most recently tested in \cite{Arya:2018sgw} with good
agreement with Planck data.

The ultimate goal in building inflation models is to have them consistent
with observational data and also be theoretically consistent.
The latter has many levels of criteria.  In \cite{Berera:2004vm}, based on
general properties of quantum field theory, conditions were listed for
theoretically consistent warm inflation models. In fact a subset
of those conditions would also apply to cold inflation models.
Amongst the criteria stated were that $\phi < M_{P}$ and 
$m_{\phi} > H$.  The former was based on general consideration
of low-energy effective field theory.  The latter is so
that inflaton particles remained sub-Hubble scale and is
basically the $\eta$-condition.   However in contrast to
cold inflation, where this condition is something one needs
to model build around, in warm inflation due to
the presence of dissipation this condition can hold
and inflation can still occur.
In Sect.~(\ref{Results})  explicit examples of warm inflation
models are shown where $\phi < M_{p}$ and $m_{\phi} > H$,
such as in Fig.~\ref{fig:phi_mphi_Qstar}.
These are
very stringent requirements for an inflation model and no
cold inflation models can achieve both of them.
The swampland criteria are basically contained within these
criteria.  In this paper we have found warm inflation models
which are consistent with all the criteria stated in \cite{Berera:2004vm}
and so in turn are also consistent with the swampland criteria. Supergravity corrections will also, in principle, be under control in such supersymmetric scenarios with sub-Planckian field values, although scenarios with $\phi> M_p$ may also be viable in supergravity with non-canonical choices of the K\"ahler potential (see e.g.~\cite{Kallosh:2010ug}).

The distributed mass model of warm inflation studied in
this paper develops the original studies to include different field- and temperature-dependent mass distributions and makes accurate comparisons with the most recent Planck data. This model fits naturally with the string landscape picture,
with different mass distributions in the model ultimately
arising from different
possible string vacua.  We found for example for
the $\Upsilon \sim \phi$ model with $\lambda \phi^4$ potential, 
full consistency
with the conditions in \cite{Berera:2004vm} and therefore
also consistency with the swampland criteria.
As the $\lambda \phi^4$ potential in many respects is
the benchmark inflation model, this consistency is
a very promising result both for warm inflation
and the DM implementation of landscape phenomenology.
We also examined other types of mass distribution functions
with varied level of success. Overall this shows
these models have a robust range of possibility.
These results encourage further understanding of DM models
in particular in the context of string theory building
on the ideas developed in \cite{Berera:1999wt}.

If one accepts string theory and thus its landscape property as the
fundamental description of nature, then it forces a rethink on
the relevance of simplicity for an inflation model. Any
point on the landscape
that leads to a consistent effective field
theory model which agrees with observational data
is just as good as any other.   We refrain from calling
this what it is, but that is the path this line of reasoning
evidently takes us down.
From the present vantage point
with so many possible vacua, it would seem easy
to argue that any working low-energy effective model
leading to observationally consistent inflation can be found
somewhere in this vast landscape.  Alternatively said, it
is hard to argue that simple low-energy models are the
only ones that could arise from this landscape.  It is
possible that future work in string theory will reduce the number of
viable vacua, but given the starting point, that seems a difficult task.
From another viewpoint, the loss of predictivity that
comes with the landscape property, equally 
encourages objective thinking
whether string theory is the best way to approach
the ultraviolet completion problem.

In a certain way of looking at it, DM models are rather complicated
in that there are many fields and several conditions to examine and
calculate.  However from another perspective they are also simple in that
they are renormalizable models, with canonical kinetic terms and
require no assumptions of the coupling of gravity.  Due
to these properties, they are very amenable to be part of
an extension of the Standard Model. All
said, from the landscape perspective, DM models are
viable models that are interesting to further explore.


\acknowledgments
RHJ acknowledges CONACyT for financial support.  AB is supported by STFC. MBG is partially supported by MINECO grant FIS2016-7819-P, and Junta de Andaluc\'ia Project FQM-101. JGR is supported by the FCT Investigator Grant No.~IF/01597/2015 and partially by the H2020-MSCA-RISE-2015 Grant No. StronGrHEP-690904 and by the CIDMA Project No.~UID/MAT/04106/2019.


\appendix

\section{Dissipative coefficient in the high temperature regime: pole approximation}\label{appendix a}
The leading contribution to the dissipative coefficient from the scalar $\chi_{i}$ fields has the following form \cite{BasteroGil:2010pb,BasteroGil:2012cm}:
\begin{equation}\label{appendix-dissipation-full}
\Upsilon=\sum_{i=1}^{N_{M}}\frac{4}{T}\left(\frac{g^{2}}{2}\right)^{2}\left(\phi-M_{i}\right)^{2}\int \frac{d^{4}p}{(2\pi)^{4}}\rho_{\chi}^{2} n_{B}(1+n_{B}) \,,
\end{equation} 
where $n_{B}=[e^{p_{0}/T}-1]^{-1}$ is the Bose-Einstein distribution and $\rho_{\chi}$ is the spectral function for the $\chi_{i}$ field. The spectral function for the $\chi_{i}$ field entering in eq.(\ref{appendix-dissipation-full}) corresponds to the fully dressed propagator, including the effect of its finite decay width into $\sigma_i$ and $\psi_{\sigma_i}$ particles:
\begin{equation}\label{appendix-spectral}
\rho_{\chi}(p,p_{0})=\frac{4\omega_{p}\Gamma_{\chi_i}}{(p_{0}^{2}-\omega_{p}^{2})^{2}+4\omega_{p}^{2}\Gamma_{\chi_i}^{2}} \,,
\end{equation}
where $\omega_{p}=\sqrt{\tilde{m}_{\chi\,i}^{2}+p^{2}}$ and $\tilde{m}_{\chi\,i}^{2}=m_{\chi\,i}^{2}+g^{2}/12+h^{2}T^{2}/8$. The leading contribution to the decay width of the $\chi_{i}$ fields is then the two-body decay $\chi_{i}\rightarrow \sigma_i\sigma_i$ and $\chi_{i}\rightarrow \bar{\psi}_{\sigma_i}\psi_{\sigma_i}$, where at finite temperature we include contributions from both decays and inverse decays, as well as thermal scattering off particles in the radiation bath. The first decay process has been computed in \cite{BasteroGil:2010pb,BasteroGil:2012cm} from the imaginary part of the $\chi_{i}$ self energy at one-loop order, yielding:
\begin{equation}\label{appendix-gamma-chi-sigma-sigma}
\Gamma_{\chi_i}(\chi_{i}\rightarrow \sigma_i\sigma_i)=\frac{h^{2}}{32\pi}\frac{m_{\chi\,i}^{2}}{\omega_{p}}F_{T}^{\sigma_i\sigma_i}(p,p_{0}) \,,
\end{equation}
where
\begin{eqnarray}
F_{T}^{\sigma_i\sigma_i}(p,p_{0}) &=& \left[ \frac{\omega_{+}^{\sigma_i}-\omega_{-}^{\sigma_i}}{p} +\frac{T}{p}\ln\left(\frac{1-e^{-\frac{\omega_{+}^{\sigma_i}}{T}}}{1-e^{-\frac{\omega_{-}^{\sigma_i}}{T}}}  \frac{1-e^{-\frac{p_{0}-\omega_{-}^{\sigma_i}}{T}}}{1-e^{-\frac{p_{0}-\omega_{+}^{\sigma_i}}{T}}} \right)  \right]\theta(p_{0}^{2}-p^{2}-4m_{\sigma_i}^{2}) \nonumber \\
&+& \left[\frac{T}{p}\ln\left(\frac{1-e^{-\frac{\omega_{+}^{\sigma_i}}{T}}}{1-e^{-\frac{\omega_{-}^{\sigma_i}}{T}}}  \frac{1-e^{-\frac{p_{0}+\omega_{-}^{\sigma_i}}{T}}}{1-e^{-\frac{p_{0}+\omega_{+}^{\sigma_i}}{T}}} \right)  \right]\theta(-p_{0}^{2}+p^{2}) \,,
\end{eqnarray}
with $m_{\sigma_i}$ denoting the scalar $\sigma_i$ mass, $\theta(x)$ the Heaviside function and 
\begin{equation}
\omega_{\pm}^{\sigma_i}=\sqrt{(k_{\pm}^{\sigma_i})^{2}+m_{\sigma_i}^{2}} \,, \qquad k_{\pm}^{\sigma_i}=\frac{1}{2} \left| p \pm p_{0}\left(1-\frac{4m_{\sigma_i}^{2}}{p_{0}^{2}-p^{2}} \right)^{1/2} \right| \,.
\end{equation}
On the other hand, the leading contribution to the decay width of the $\chi_{i}$ fields into two fermions $\bar{\psi}_{\sigma_i}\psi_{\sigma_i}$, has been computed in \cite{BasteroGil:2010pb} from the imaginary part of the $\chi_{i}$ self energy at one-loop order, yielding:
\begin{equation}\label{appendix-gamma-chi-psisigma-psisigma}
\Gamma_{\chi_i}(\chi_{i}\rightarrow \bar{\psi}_{\sigma_i}\psi_{\sigma_i})=\frac{h^{2}}{32\pi\omega_{p}}F_{T}^{\bar{\psi}_{\sigma_i}\psi_{\sigma_i}}(p,p_{0}) \,,
\end{equation}
where
\begin{eqnarray}
F_{T}^{\bar{\psi}_{\sigma_i}\psi_{\sigma_i}}(p,p_{0}) &=& (p_{0}^{2}-p^{2})\left[1-\frac{4m_{\psi_{\sigma_i}}^{2}}{p_{0}^{2}-p^{2}}\right]\left[ \frac{\omega_{+}^{\psi_{\sigma_i}}-\omega_{-}^{\psi_{\sigma_i}}}{p} +\frac{T}{p}\ln\left(\frac{1+e^{-\frac{\omega_{+}^{\psi_{\sigma_i}}}{T}}}{1+e^{-\frac{\omega_{-}^{\psi_{\sigma_i}}}{T}}}  \frac{1+e^{-\frac{p_{0}-\omega_{-}^{\psi_{\sigma_i}}}{T}}}{1+e^{-\frac{p_{0}-\omega_{+}^{\psi_{\sigma_i}}}{T}}} \right)  \right]\theta(p_{0}^{2}-p^{2}-4m_{\psi_{\sigma_i}}^{2}) \nonumber \\
&& + (p_{0}^{2}-p^{2})\left[1-\frac{4m_{\psi_{\sigma_i}}^{2}}{p_{0}^{2}-p^{2}}\right]\frac{T}{p}\ln\left(\frac{1+e^{-\frac{\omega_{+}^{\psi_{\sigma_i}}}{T}}}{1+e^{-\frac{\omega_{-}^{\psi_{\sigma_i}}}{T}}} \frac{1+e^{-\frac{p_{0}+\omega_{-}^{\psi_{\sigma_i}}}{T}}}{1+e^{-\frac{p_{0}+\omega_{+}^{\psi_{\sigma_i}}}{T}}} \right)\theta(-p_{0}^{2}+p^{2}) \,,
\end{eqnarray}
then $m_{\psi_{\sigma_i}}$ is the fermionic $\psi_{\sigma_i}$ mass and 
\begin{equation}
\omega_{\pm}^{\psi_{\sigma_i}}=\sqrt{(k_{\pm}^{\psi_{\sigma_i}})^{2}+m_{\psi_{\sigma_i}}^{2}} \,, \qquad k_{\pm}^{\psi_{\sigma_i}}=\frac{1}{2} \left| p \pm p_{0}\left(1-\frac{4m_{\psi_{\sigma_i}}^{2}}{p_{0}^{2}-p^{2}} \right)^{1/2} \right| \,.
\end{equation}
An analysis of Eq.(\ref{appendix-dissipation-full}) indicates that the behaviour of the dissipation coefficient in the different temperature and interaction regimes is determined by an interplay between the spectral function and the thermal occupation numbers. When the temperature is large, the occupation numbers are also large, so in this regime, the poles of the spectral functions will dominate the integral and this will be referred to as the pole approximation. As such, the pole approximation works well in the high-$T$ region. The dissipation coefficient about its pole at $p_{0}\simeq \omega_{p}$ is:
\begin{equation}\label{appendix-upsilon-pole-approximation}
\Upsilon =\sum_{i=1}^{N_{M}}\frac{2}{T}\left(\frac{g^{2}}{2}\right)^{2}\left(\phi-M_{i}\right)^{2}\int\frac{d^{3}p}{(2\pi)^{3}}\frac{n_{B}(1+n_{B})}{\Gamma_{\chi_i}\omega_{p}^{2}} \,.
\end{equation}
For on-shell $\chi_{i}$ modes, decays into light scalars and fermions are equally probable, with partial decay widths given by Eqs.~(\ref{appendix-gamma-chi-sigma-sigma}) and (\ref{appendix-gamma-chi-psisigma-psisigma}). By treating the scalar $\sigma_i$ and the fermion $\psi_{\sigma_i}$ as light fields, we evaluate both decay rates in the limit $m_{\sigma_i}\ll T$ ($m_{\psi_{\sigma_i}}\ll T$), so $\omega_{\pm}^{\sigma_i}/T\simeq k_{\pm}^{\sigma_i}/T$ ($\omega_{\pm}^{\psi_{\sigma_i}}/T\simeq k_{\pm}^{\psi_{\sigma_i}}/T$); this estimation also yields that $k_{\pm}^{\sigma_i}=k_{\pm}^{\psi_{\sigma_i}}=\mid p_{0} \pm p \mid/2$. Finally, in the limit $m_{\chi_i}\ll T$, we have that both partial decay widths are given by: 
\begin{eqnarray}
\Gamma_{\chi_i}(\chi_{i}\rightarrow\sigma_i\sigma_i) &=& \frac{h^{2}}{8\pi}\frac{m^{2}_{\chi_i}}{\tilde{m}_{\chi_i}}\frac{T}{\omega_{p}} \,, \label{appendix-gamma-chi-sigma-sigma-approx} \\
\Gamma_{\chi_i}(\chi_i\rightarrow \bar{\psi}_{\sigma_i}\psi_{\sigma_i}) &=& \frac{h^{2}}{128\pi}\frac{\tilde{m}^{3}_{\chi_i}}{\omega_{p}T} \,, \label{appendix-gamma-chi-psisigma-psisigma-approx}
\end{eqnarray}
yielding a total decay rate:
\begin{equation}\label{appendix-full-Gamma}
\Gamma_{\chi_i}=\frac{h^{2}}{128\pi}\frac{T\tilde{m}_{\chi_{i}}}{\omega_{\chi_{i}}(p)}\left[16\frac{m_{\chi_{i}}^{2}}{\tilde{m}_{\chi_{i}}^{2}}+\frac{\tilde{m}_{\chi i}^{2}}{T^{2}}\right] \,.
\end{equation}
Finally we compute the integral in Eq.~(\ref{appendix-upsilon-pole-approximation}), by following the same procedure as in \cite{Berera:1998px}, obtaining the total scalar dissipative coefficient:
\begin{equation}\label{appendix-Upsilon-highT}
\Upsilon^{S}(\phi,T)= \sum_{i=1}^{N_{M}}\frac{32 g^{4}}{\pi h^{2}\left[16\frac{m_{\chi_{i}}^{2}}{\tilde{m}_{\chi_{i}}^{2}}+\frac{\tilde{m}_{\chi i}^{2}}{T^{2}}\right]}\ln\left(\frac{2T}{\tilde{m}_{\chi_{i}}}\right)\frac{\left(\phi-M_{i}\right)^{2}}{\tilde{m}_{\chi_{i}}} \,.
\end{equation}

\section{Scalar decay width average}\label{appendix b}
The thermal average of the scalar decay width is given by: 
\begin{equation}\label{decay-width-definition-appendix}
\bar{\Gamma}_{\chi_i}=\frac{1}{n_{B}}\int\frac{d^{3}p}{(2\pi)^{3}}\Gamma_{\chi_i}f_{B} \,,
\end{equation}
where $f_{B}(\mathbf{p}/T)$ is the Bose-Einstein distribution and $n_{B}(\tilde{m}_{\chi_{i}}/T)$ the associated number density. Hence we have: 
\begin{equation}\label{decay-width-1-appendix}
\bar{\Gamma}_{\chi_i}=\frac{h^{2}T\tilde{m}_{\chi_{i}}}{128\pi}\left[16\frac{m_{\chi_{i}}^{2}}{\tilde{m}_{\chi_{i}}^{2}}+\frac{\tilde{m}_{\chi i}^{2}}{T^{2}}\right]I(\tilde{m}_{\chi_{i}}/T) \,, \quad I(\tilde{m}_{\chi_{i}}/T) =  \frac{\int\frac{d^{3}p}{(2\pi)^{3}}\frac{f_{B}}{\omega_{\chi_{i}}(p)}}{\int\frac{d^{3}p}{(2\pi)^{3}}f_{B}} \,.
\end{equation}
The integral factor $I(\tilde{m}_{\chi_{i}}/T)$ can be obtained numerically and is well approximated by: 
\begin{equation}\label{integral-average-decay-width-appendix}
I(\tilde{m}_{\chi_{i}}/T) = \frac{1}{T}\frac{\int\frac{dxx^{2}}{\sqrt{x^{2}+a^{2}}}(e^{\sqrt{x^{2}+a^{2}}}-1)^{-1}}{\int dxx^{2}(e^{\sqrt{x^{2}+a^{2}}}-1)^{-1}} \simeq \frac{1}{T}\frac{0.68}{(1+0.77a)} \,,
\end{equation}
where we have defined $x=p/T$ and $a=\tilde{m}_{\chi_{i}}/T$. Hence the thermal average of the scalar decay width is: 
\begin{equation}\label{average-decay-rate-approx1-appendix}
\bar{\Gamma}_{\chi_i}\simeq \frac{h^{2}\tilde{m}_{\chi_{i}}}{128\pi}\left[16\frac{m_{\chi_{i}}^{2}}{\tilde{m}_{\chi_{i}}^{2}}+\frac{\tilde{m}_{\chi i}^{2}}{T^{2}}\right]\frac{0.68}{(1+0.77 f^{1/2})} \,. 
\end{equation}
Moreover, in the limit $m_{\chi_i}/T\lesssim1$ we have that the effective finite thermal mass $\tilde{m}_{\chi_{i}}$ can be fittingly taken as $\tilde{m}_{\chi_{i}}\simeq f^{1/2}T$, where  $f=f(g,h)=g^{2}/12+h^{2}/8$. Therefore the average decay width becomes: 
\begin{equation}\label{average-decay-rate-approx2-appendix}
\bar{\Gamma}_{\chi_i}\simeq \frac{h^{2}f^{1/2}}{128\pi}\left[\frac{16}{f}\frac{m_{\chi_{i}}^{2}}{T^{2}}+f\right]\frac{0.68}{(1+0.77 f^{1/2})}T \,. 
\end{equation}

\section{Effective potential at finite temperature }\label{appendix c}
We start by considering the contribution of the light fermions in the tower to the finite temperature effective potential, which is given by \cite{Dolan:1973qd,Kapusta:2006,Cline:1996mga}:
\begin{equation}\label{VT-fermions-C1}
V_{T}^{\psi_{\chi_i}} \simeq \frac{1}{2}\sum_{i}\left\{ -\frac{7\pi^{2}}{180}T^{4}+\frac{\tilde{m}_{\psi_{\chi_i}}^{2}T^{2}}{12}+\frac{\tilde{m}_{\psi_{\chi_i}}^{4}}{16\pi^{2}}\left[\ln\left(\frac{\mu^{2}}{T^{2}}\right)-c_{f}\right]   \right\}\,,
\end{equation}
where $\mu$ is the $\overline{\text{MS}}$ renormalization scale, $c_{f}=2.635$, and the effective thermal fermion masses $\tilde{m}_{\psi i}^{2}=g^{2}(\phi-M_{i})^{2}+h^{2}T^{2}/8$. Note that the overall factor $1/2$ in front of the sum is related to the Majorana nature of the fermions in the SUSY model. In the continuum limit we may write this in the form:
\begin{eqnarray}
V_{T}^{\psi_{i}} &\simeq& \frac{T^{4}}{2}\sum_{i}\left\{ -\frac{7\pi^{2}}{180}+\left[\frac{g^{2}(\phi-M_{i})^{2}}{T^{2}}+\frac{h^{2}}{8}\right]\frac{1}{12} \right. \nonumber\\
&& \qquad\left. +\left[\frac{g^{4}(\phi-M_{i})^{4}}{T^{4}}+\frac{h^{2}}{4}\frac{g^{2}(\phi-M_{i})^{2}}{T^{2}}+\frac{h^{4}}{64}\right]\frac{1}{16\pi^{2}}\left[\ln\left(\frac{\mu^{2}}{T^{2}}\right)-c_{f}\right] \right\} \nonumber\\
&\simeq& \frac{T^{4}}{2}\left\{\left[-\frac{7\pi^{2}}{180}+\frac{h^{2}}{96}+\frac{h^{4}}{1024\pi^{2}}\left[\ln\left(\frac{\mu^{2}}{T^{2}}\right)-c_{f}\right]\right]\int_{M_{-}}^{M_{+}}dMn(M) \right. \nonumber\\
&& \qquad\left. +\frac{g^{2}}{T^{2}}\left[\frac{1}{12}+\frac{h^{2}}{64\pi^{2}}\left[\ln\left(\frac{\mu^{2}}{T^{2}}\right)-c_{f}\right]\right]\int_{M_{-}}^{M_{+}}dMn(M)(\phi-M)^{2}\right. \nonumber\\
&& \qquad\left. +\frac{1}{16\pi^{2}}\frac{g^{4}}{T^{4}}\left[\ln\left(\frac{\mu^{2}}{T^{2}}\right)-c_{f}\right]\int_{M_{-}}^{M_{+}}dMn(M)(\phi-M)^{4} \right\} \nonumber\\
&\simeq& \frac{T^{4}}{2}\left\{\left[-\frac{7\pi^{2}}{180}+\frac{h^{2}}{96}+\frac{h^{4}}{1024\pi^{2}}\left[\ln\left(\frac{\mu^{2}}{T^{2}}\right)-c_{f}\right]\right]\left[\frac{2T}{g}n(\phi)+\cdots\right]\right. \nonumber\\
&& \qquad\left. +\frac{g^{2}}{T^{2}}\left[\frac{1}{12}+\frac{h^{2}}{64\pi^{2}}\left[\ln\left(\frac{\mu^{2}}{T^{2}}\right)-c_{f}\right]\right]\left[\frac{2}{3}\frac{T^{3}}{g^{3}}n(\phi)+\cdots\right] \right. \nonumber\\
&& \qquad\left. +\frac{1}{16\pi^{2}}\frac{g^{4}}{T^{4}}\left[\ln\left(\frac{\mu^{2}}{T^{2}}\right)-c_{f}\right]\left[\frac{2}{5}\frac{T^{5}}{g^{5}}n(\phi)+\cdots\right] \right\} \nonumber\\
&\simeq& \frac{T^{5}}{2g}n(\phi)\left\{-\frac{7\pi^{2}}{90}+\frac{1}{18}+\frac{h^{2}}{48}+\frac{1}{16\pi^{2}}\left(\frac{2}{5}+\frac{h^{2}}{6}+\frac{h^{4}}{32}\right)\left[\ln\left(\frac{\mu^{2}}{T^{2}}\right)-c_{f}\right]\right\}+\cdots \label{VT-fermions-continuum}\,,
\end{eqnarray}
where the integrals are in general
\begin{eqnarray}
\int_{M_{-}}^{M_{+}}dMn(M)(\phi-M)^{n} &=& \int_{-T/g}^{T/g}dx\left[n(\phi)-xn'(\phi)+\frac{x^{2}}{2!}n''(\phi)+\cdots  \right]x^{n} \nonumber\\
&=& n(\phi)\frac{x^{n+1}}{n+1}-n'(\phi)\frac{x^{n+2}}{n+2}+n''(\phi)\frac{x^{n+3}}{n+3}+\cdots\Bigr|_{-T/g}^{T/g} \nonumber\\
&=& 
\left\{
\begin{array}{c}
\frac{2}{n+1}\left(\frac{T}{g}\right)^{n+1}n(\phi)+\cdots    \qquad\,, n=0,2,4,\ldots  \\
-\frac{2}{n+2}\left(\frac{T}{g}\right)^{n+2}n'(\phi)+\cdots    \qquad\,, n=1,3,5,\ldots 
\end{array}
\right. 
\end{eqnarray}
We have considered the local approximation, i.e. that only states in the vicinity of $M=\phi(t)$ are light at any given time. Note that to obtain the derivatives of this potential correction with respect to the field, one needs to take into account that both the integrand and the integration limits are $\phi$-dependent, the end result corresponding to differentiating eq. (\ref{VT-fermions-continuum}), such that:
\begin{eqnarray}
V_{T,\phi}^{\psi_{i}} &\simeq& \frac{T^{5}}{2g}n'(\phi)\left\{-\frac{7\pi^{2}}{90}+\frac{1}{18}+\frac{h^{2}}{48}+\frac{1}{16\pi^{2}}\left(\frac{2}{5}+\frac{h^{2}}{6}+\frac{h^{4}}{32}\right)\left[\ln\left(\frac{\mu^{2}}{T^{2}}\right)-c_{f}\right]\right\}+\cdots \label{dVTdphi-fermions-continuum}\,, \\
V_{T,\phi\phi}^{\psi_{i}} &\simeq& \frac{T^{5}}{2g}n''(\phi)\left\{-\frac{7\pi^{2}}{90}+\frac{1}{18}+\frac{h^{2}}{48}+\frac{1}{16\pi^{2}}\left(\frac{2}{5}+\frac{h^{2}}{6}+\frac{h^{4}}{32}\right)\left[\ln\left(\frac{\mu^{2}}{T^{2}}\right)-c_{f}\right]\right\}+\cdots \label{d2VTd2phi-fermions-continuum}\,.
\end{eqnarray}
Nevertheless, in the local approximation the density of states could be taken only as temperature-dependent $n\sim T^{-1}$, and independent of the field; therefore $n'(\phi)=n''(\phi)=0$. Thus we have $N_M\simeq 2Tn/g$, hence $n\simeq gN_M/(2T)$; but most importantly $V_{T,\phi}^{\psi_{\chi_i}}=V_{T,\phi\phi}^{\psi_{i}}\simeq 0$. Finally the finite temperature corrections to the effective potential due to the fermionic tower, in the local approximation at leading order, can be written as:
\begin{equation}\label{VT-fermions-continuum-uniform}
V_{T}^{\psi_{i}} \simeq  \frac{N_{M}}{4}T^{4}\left\{-\frac{7\pi^{2}}{90}+\frac{1}{18}+\frac{h^{2}}{48}+\frac{1}{16\pi^{2}}\left(\frac{2}{5}+\frac{h^{2}}{6}+\frac{h^{4}}{32}\right)\left[\ln\left(\frac{\mu^{2}}{T^{2}}\right)-c_{f}\right]\right\}\,. 
\end{equation}
Next we examine the contribution of the light bosons in the tower to the finite temperature effective potential given by \cite{Dolan:1973qd,Kapusta:2006,Cline:1996mga}:
\begin{equation}\label{VT-scalars-C1}
V_{T}^{\chi_{i}} \simeq 2\sum_{i}\left\{ -\frac{\pi^{2}}{90}T^{4}+\frac{\tilde{m}_{\chi_i}^{2}T^{2}}{24}-\frac{\tilde{m}_{\chi_i}^{3}T}{12\pi}-\frac{\tilde{m}_{\chi_i}^{4}}{64\pi^{2}}\left[\ln\left(\frac{\mu^{2}}{T^{2}}\right)-c_{b}\right] \right\}\,,
\end{equation}
where again $\mu$ denotes the $\overline{\text{MS}}$ renormalization scale, $c_{b}=5.41$, and the effective thermal boson masses $\tilde{m}_{\chi_i}^{2}=g^{2}(\phi-M_{i})^{2}+g^{2}T^{2}/12+h^{2}T^{2}/8$. Note the overall factor 2 in front of the sum, which represents the fact that the $\chi_{i}$'s scalars are complex fields. Following the same procedure as for the fermionic tower, we have for the bosonic sector: 
\begin{eqnarray}
V_{T}^{\chi_{i}} &\simeq& \frac{2T^{5}}{g}n(\phi)\left\{-\frac{\pi^{2}}{45}+\frac{f^{2}(g,h)}{16\pi}\ln\left(\frac{\sqrt{f(g,h)}}{\sqrt{1+f(g,h)}+1}\right)+\frac{f(g,h)}{12} -\frac{\left(2+5f(g,h)\right)}{48\pi}\sqrt{1+f(g,h)}\right.\nonumber\\
&& \qquad\qquad\left.-\frac{1}{16\pi^{2}}\left(\frac{1}{10}+\frac{f(g,h)}{3}+\frac{f^{2}(g,h)}{2}\right)\left[\ln\left(\frac{\mu^{2}}{T^{2}}\right)-c_{b}\right]\right\}+\cdots \,,\label{VT-bosons-continuum}
\end{eqnarray}
where $f(g,h)=g^{2}/12+h^{2}/8$. In the local approximation $n(\phi)\sim T^{-1}$, hence $n(\phi)\simeq gN_{\chi}/(2T)$, and the inflaton $\phi$ derivatives are zero: $V_{T,\phi}^{\chi_{i}}=V_{T,\phi\phi}^{\chi_{i}}\simeq 0$. Finally the finite temperature corrections to the effective potential due to the bosonic tower, in the local approximation at leading order, can be written in the form:
\begin{eqnarray}
V_{T}^{\chi_{i}} &\simeq& N_MT^{4}\left\{-\frac{\pi^{2}}{45}+\frac{f^{2}(g,h)}{16\pi}\ln\left(\frac{\sqrt{f(g,h)}}{\sqrt{1+f(g,h)}+1}\right)+\frac{f(g,h)}{12} -\frac{\left(2+5f(g,h)\right)}{48\pi}\sqrt{1+f(g,h)}\right.\nonumber\\
&& \qquad\qquad\left.-\frac{1}{16\pi^{2}}\left(\frac{1}{10}+\frac{f(g,h)}{3}+\frac{f^{2}(g,h)}{2}\right)\left[\ln\left(\frac{\mu^{2}}{T^{2}}\right)-c_{b}\right]\right\} \,.\label{VT-bosons-continuum-uniform}
\end{eqnarray}


\end{document}